\documentclass[conference]{IEEEtran}

\usepackage{xspace}
\usepackage{algorithm}
\usepackage{algorithmic}
\usepackage{tabularx}
\usepackage{graphicx}
\usepackage{subcaption}
\usepackage{amsmath}
\usepackage{amsfonts}
\usepackage{url}
\usepackage{makecell}
\usepackage{orcidlink}
\usepackage{amssymb}

\hypersetup{
  colorlinks=false,
  linkbordercolor=white,
 urlbordercolor=white,
pdfborder={0 0 0}
}

\usepackage{xspace}

\newcommand{\Prismo}{{\textsc{\small{Prismo}}}\xspace}

\newcommand{\shortsectionBf}[1]{\vspace{2pt}
\noindent {\bf #1}}

\begin{document}


\title{Prismo: A Decision Support System for Privacy-Preserving ML Framework Selection}

\author{
    Nges Brian Njungle\orcidlink{0009-0006-3393-6851}$^{1}$,
    Eric Jahns\orcidlink{0009-0004-5511-7975}$^{1}$,
    Luigi Mastromauro\orcidlink{0009-0004-5511-7975}$^{1}$, 
    Edwin P. Kayang\orcidlink{0009-0003-7158-5608}$^{1}$ \\
    Milan Stojkov\orcidlink{0000-0002-0602-0606}$^{2}$,
    and Michel A. Kinsy\orcidlink{0000-0002-1432-6939}$^{1}$\\
    $^{1}$STAM Center, Ira A. Fulton Schools of Engineering, Arizona State University, Tempe, AZ 85281, USA\\
    $^{2}$Faculty of Technical Sciences, University of Novi Sad, Novi Sad, Serbia\\
    Corresponding Author: nnjungle@asu.edu
}

\maketitle

\begin{abstract}
Machine learning has become a crucial part of our lives, with applications spanning nearly every aspect of our daily activities. However, using personal information in machine learning applications has sparked significant security and privacy concerns about user data. 
To address these challenges, different privacy-preserving machine learning (PPML) frameworks have been developed to protect sensitive information in machine learning applications. These frameworks generally attempt to balance design trade-offs such as computational efficiency, communication overhead, security guarantees, and scalability. Despite the advancements, selecting the optimal framework and parameters for specific deployment scenarios remains a complex and critical challenge for privacy and security application  developers.

We present \Prismo, an open-source recommendation system designed to aid in selecting optimal parameters and frameworks for different  PPML application scenarios. \Prismo enables users to explore a comprehensive space of PPML frameworks through various properties based on user-defined objectives. It supports automated filtering of suitable candidate frameworks by considering parameters such as the number of parties in multi-party computation or federated learning and computation cost constraints in homomorphic encryption. \Prismo models every use case into a Linear Integer Programming optimization problem, ensuring tailored solutions are recommended for each scenario. 
We evaluate \Prismo's effectiveness through multiple use cases, demonstrating its ability to deliver best-fit solutions in different deployment scenarios.
\end{abstract}


\maketitle

\section{Introduction}
\label{sec:introduction}

Machine learning (ML) profoundly impacts various sectors, including healthcare, finance, education, and cybersecurity, revolutionizing how industries operate and make decisions \cite{applicationofml}. In healthcare specifically, ML is utilized for various applications, such as diagnosing diseases, predicting health outcomes, aiding in drug development, providing treatment recommendations, enabling continuous health monitoring, and analyzing genetic data. \cite{ml_apps}.
This transformative potential is not just confined to enhancing efficiencies or improving existing systems. Still, it is poised to revolutionize entire industries and redefine how services are delivered globally.
As ML continues to play an integral role in today's society, its reliance on large datasets (often aggregated from different sources) raises concerns about data security and privacy. 
The conflicting interests of data providers and ML model developers heighten these privacy concerns. Data providers focus on protecting sensitive information, while model developers are equally invested in protecting their proprietary algorithms and models. The stakes are even higher in high-risk areas such as healthcare, where privacy breaches can have life-altering consequences. This growing tension has underscored the urgent need for solutions that balance data privacy with the effective deployment of ML, ensuring that confidentiality is maintained while enabling collaboration within the field.

Privacy-Preserving Machine Learning (PPML) provides solutions for protecting both customer data and ML models while facilitating collaboration and productivity with ML \cite{xu2021privacypreserving}. By implementing PPML, organizations can leverage ML's full potential without compromising sensitive information.
Several privacy-preserving techniques have emerged, including federated learning, secure multi-party computation, differential privacy, trusted execution environments, and homomorphic encryption. 
These techniques enable secure collaboration in ML while protecting data and models, helping shape the future of privacy-conscious ML applications \cite{typesofppml}.
Various tools, such as protocols and libraries, are typically available to implement these techniques. Over time, a range of frameworks has been developed and publicly released that support the integration of these techniques into ML systems. Yet, each framework has strengths, limitations, and varying performance across application scenarios and environments.

One of the key challenges in adopting PPML is determining how to compare the proposed frameworks, select the most appropriate ones, and rank them for diverse use cases. While numerous surveys and systematizations of knowledge aim to address a similar issue, their contributions are often limited by the absence of comprehensive and quantitative metrics that enable rigorous, scenario-specific comparisons. This shortcoming becomes particularly pronounced when frameworks are based on fundamentally different underlying techniques, as existing studies frequently lack the flexibility required to accommodate the diverse requirements and constraints of real-world PPML deployment scenarios.

In this work, we introduce \Prismo, a recommender system designed to identify the most suitable PPML frameworks for any given deployment scenario through a multi-objective linear integer programming optimization approach. 
To facilitate a comprehensive analysis of PPML frameworks, we study seventy-four PPML frameworks across different techniques. This study established a baseline to identify and establish a set of detailed and common criteria for capturing the most critical features needed to understand their fitness in different application scenarios. 
Our analysis criteria incorporate essential aspects common to all PPML frameworks, such as threat models, training and inference support, framework accessibility, and the performance of the framework validated across different use cases.
By applying this method, we collect data from the seventy-four frameworks. The data is then imported into the \Prismo's database. \Prismo transforms these data into quantitative values, which are then used in solving linear integer programming problems to make recommendations tailored towards every specific use case. 
\Prismo offers a practical tool that delivers effective recommendations for deploying PPML solutions, ensuring security and privacy considerations are met while identifying the optimal solution for each unique deployment scenario.
We chose the multi-objective linear integer programming optimization approach to solve this problem because it gives us the most efficient way to comprehensively and quantitatively represent each PPML solution, compare these solutions efficiently, and also offers the flexibility needed to effectively represent different deployment scenarios.
To evaluate \Prismo, we considered three use cases and compared their recommended outputs with other frameworks also represented in \Prismo. 
Our evaluation also involves leveraging the literature on these frameworks and utilizing their key features to rank them with linear integer programming. We study the recommended frameworks to ensure that they align as the best-fitted solutions for these deployment instances. Across all analyzed cases, \Prismo consistently recommends the most suitable solution for the use case.
We employ this approach for evaluation since there are no related benchmarks for evaluating systems of this type. 
In this work, our contributions are as follows:
\begin{itemize} 
    \item We introduce \Prismo, a PPML recommender system designed for every application use case. It evaluates factors such as security, robustness, and scalability to suggest the most suitable PPML frameworks for every scenario. 

    \item We evaluate seventy-four frameworks, establishing fine-grained criteria to capture essential information across all PPML frameworks. The data captured from these frameworks provides a solid foundation to evaluate the performance of \Prismo and provide a basis for its real-world deployment and usage.
    \item  We offer an open-source repository with Dockerized implementations of various PPML frameworks and comprehensive documentation, enabling developers and researchers to integrate and use these frameworks easily.
\end{itemize}

\section{Background}
\label{sec:background}

Machine learning is often described as the science of enabling computers to perform tasks without explicitly programming them \cite{MohammadAlRubaie-2019}. It is typically divided into three primary sub-fields: Unsupervised Learning, Supervised Learning, and Reinforcement Learning, each addressing different types of problems and data structures.
Unsupervised learning algorithms handle raw, unlabeled data to uncover hidden patterns, groupings, or structures within datasets. These methods are instrumental in exploratory data analysis and dimensionality reduction, where domain knowledge is scarce or unavailable \cite{ZoubinGhahramani-2004}.
Supervised learning, in contrast, focuses on mapping input data to corresponding output labels by learning from a labeled dataset \cite{TammyJiang-2020}. This type of learning is widely used in predictive modeling tasks such as classification and regression. 
Reinforcement learning takes a unique approach by training an agent to interact with an environment and maximize a cumulative reward signal \cite{KeerthanaSivamayil-2023}. It thrives in dynamic and sequential decision-making scenarios. 

The rapid proliferation of big data has enabled the growth of a specialized branch of ML known as deep learning. Deep learning uses deep neural networks, composed of multiple layers of linear and non-linear transformations, to extract complex features from raw data \cite{IqbalSarker-2021}. This flexibility makes deep learning applicable across various domains, including unsupervised, supervised, and reinforcement learning. Notable breakthroughs in areas such as image recognition, speech processing, and natural language understanding can be attributed to architectures like convolutional neural networks (CNNs) and recurrent neural networks (RNNs) \cite{AlexKrizhevsky-2012}. Large language models such as ChatGPT and Gemini exemplify the potential of deep learning to revolutionize human-computer interaction, enabling tasks like real-time translation, automated writing assistance, and conversational AI. However, various techniques have been developed for integration during ML training and inference to address privacy and security challenges. The following sub-sections explore five widely adopted methods, highlighting their significance in protecting sensitive data and maintaining trust in ML.

\subsection{Federated Learning}
Federated Learning (FL) is a technique that facilitates collaborative model training across multiple decentralized edge devices or servers while maintaining the privacy of raw data. This decentralized and distributed approach is particularly beneficial in environments where data privacy and security are critical, since it ensures that sensitive information remains confined to the edge devices themselves \cite{10381965}. 
FL has gained prominence in applications where data is distributed and often highly sensitive  \cite{TEO2024101419, liu2023efficientsecurefederatedlearning, Nguyen_2021}. 
The FL process typically begins with a central server distributing an initial global model to all participating edge devices. Each device trains the model locally using its private dataset, generating updates based on its specific data characteristics \cite{Bharati_2022}. These updates, often in the form of weights or gradients, are then transmitted back to a central server or aggregator. The server consolidates these updates to refine and improve the global model, which is then redistributed to the devices. This iterative process continues until the global model achieves the desired level of performance \cite{10158622}.

\begin{figure}[http]
    \includegraphics[width=\linewidth]{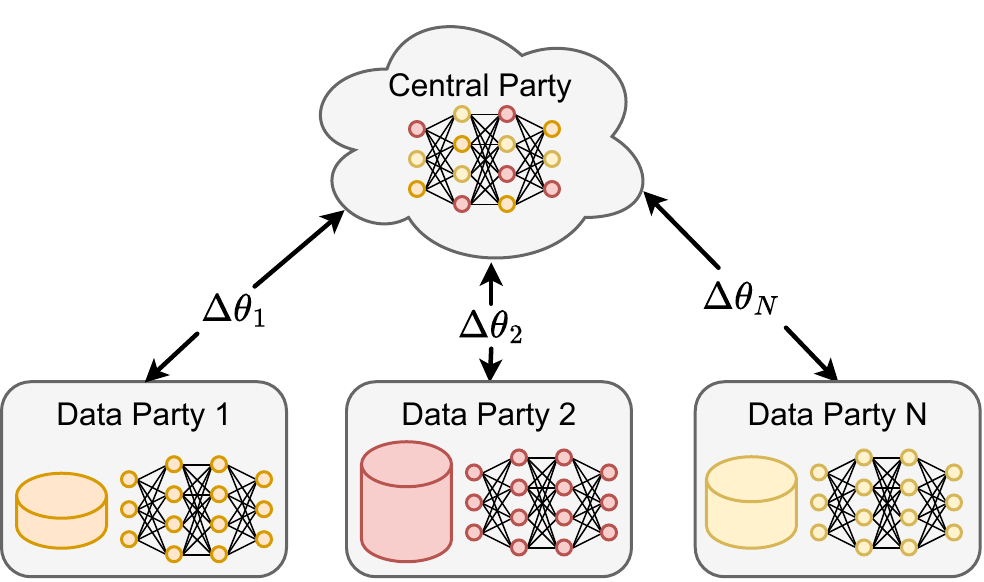}
    \captionsetup{justification=centering}
    \caption{Demonstration of centralized federated learning. Parties send locally trained weights to the centralized server, aggregating them and returning the updates to parties.}
    \label{fig:fl_illustration}
\end{figure}

One of the key strengths of FL is its ability to leverage diverse and heterogeneous datasets present on edge devices, enabling the creation of more generalized and robust models. 
Despite its advantages, FL also faces certain limitations. It does not inherently provide the strong security guarantees of cryptographic methods, leaving it vulnerable to poisoning and inference attacks \cite{fl_poisoning,fl_inference}. Furthermore, FL is best suited for collaborative scenarios where devices can communicate effectively and adhere to a common training framework. Challenges such as uneven data distribution, varying computational capacities of devices, and communication overhead can impact the performance and scalability of FL \cite{Bharati_2022}. 
Figure \ref{fig:fl_illustration} illustrates how FL operates in centralized setups. Here, edge devices send their locally trained weights ($\theta$) to a central server, which aggregates these updates ($\Delta \theta$) to refine the global model and then sends it back to all devices. 
This approach simplifies coordination but relies heavily on the central server's reliability. 
The FL frameworks analyzed for \Prismo include \cite{article,  10.5555/3546258.3546484, Beutel2020FlowerAF, garcia2022flute, he2020fedml, caldas2019leaf, DBLP:journals/corr/abs-2007-10987, WAHRSTATTER2024101174, wang2020automatedpancreassegmentationusing, galtier2019substraframeworkprivacypreservingtraceable, lai2022fedscalebenchmarkingmodelperformance, xie2022federatedscopeflexiblefederatedlearning}.

\subsection{Differential Privacy}

Differential Privacy (DP) is a PPML technique aimed at protecting sensitive information during ML training and against malicious attempts to extract private data \cite{MartinAbadi-2016}. By introducing random noise to the training data, DP obscures individual data points, ensuring that the inclusion or exclusion of any single data entry has a negligible impact on the model's output. 
This method enables organizations to derive insights from data while upholding strict privacy standards. The degree of privacy in DP is directly proportional to the amount of noise added; higher noise levels generally ensure better privacy but at the expense of model accuracy. 
This trade-off between privacy and performance is the main challenge of DP. 

\begin{figure}[http]
    \begin{center}
    \includegraphics[width=\linewidth]{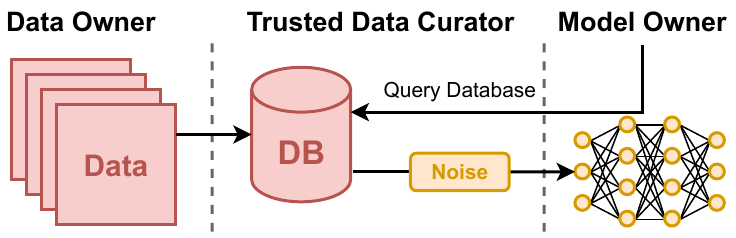}
    \captionsetup{justification=centering}
    \caption{Demonstration of differential privacy. The user adds differentially private noise through a sampling process. The obfuscated data is used to train or infer the model.}
    \label{fig:dp_illustration}
    \end{center}
\end{figure}

While it protects data confidentiality effectively, it lacks the strong security guarantees of cryptographic primitives, leaving it vulnerable to specific types of attacks, such as data poisoning or inference attacks \cite{9775757, dpattacks}.
Additionally, the noise introduced can significantly reduce accuracy, particularly in scenarios with small datasets or highly sensitive models. 
Figure \ref{fig:dp_illustration} depicts a DP setting where a party adds differentially private noise to their data before transmitting it to an untrusted model owner for training or inference. 
This approach ensures that information is protected at the source, even if the model owner is compromised. 
The DP works analyzed for \Prismo include \cite{MartinAbadi-2016, NhatHaiPhan-2017, MahawagaArachchige-2020, MaoguoGong-2020, YufengWang-2020, ZhiqiBu-2020, MdShohidulIslam-2023, TamasMadl-2023, KamilAdamczewski-2023}.

\subsection{Trusted Execution Environments}

A Trusted Execution Environment (TEE) is a secure enclave within a computing device that combines hardware and a dedicated operating system to provide a secure execution environment. 
TEEs ensure that specific processes remain isolated from others running on the device, whether in the regular operating system or within the TEE itself. This isolation protects sensitive data and computations from unauthorized access \cite{tee_intro}. 
By isolating computations and maintaining strict access controls, TEEs ensure that the processes within the TEE remain secure even if the primary operating system is compromised. However, this high level of security comes with trade-offs. TEEs are typically resource-constrained, with limited computational power and memory, which makes them expensive to deploy for resource-intensive tasks. 
Moreover, side-channel and timing attacks have been demonstrated against all currently available TEEs, raising concerns about their resilience \cite{jauernig2020trusted, munoz2023survey}.

\begin{figure}[http]
    \begin{center}
    \includegraphics[width=\linewidth]{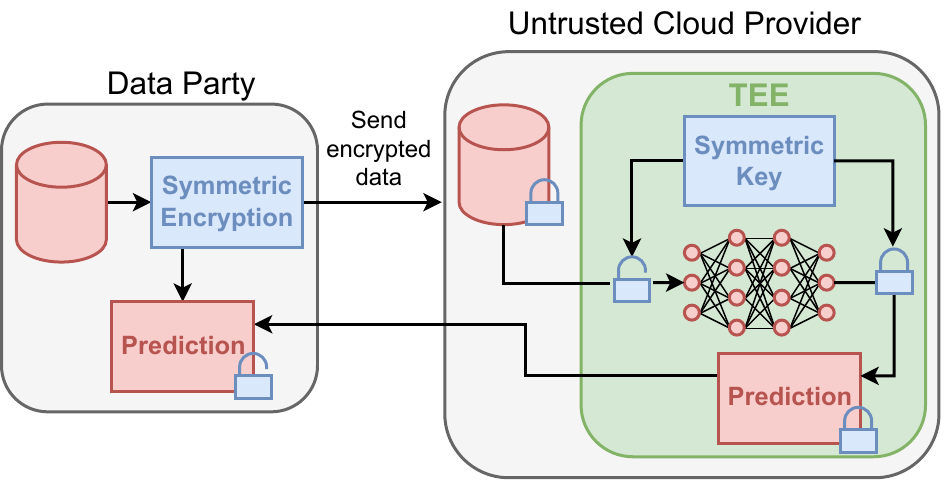}
    \captionsetup{justification=centering}
    \caption{Demonstration of a trusted execution environment. Data is encrypted using symmetric cryptography and transferred to the TEE, where it is decrypted, inferred on the model, and the results are re-encrypted and sent back to the user.}
    \label{fig:tee_illustration}
    \end{center}
\end{figure}

Figure \ref{fig:tee_illustration} illustrates a PPML scenario using a TEE for inference in a general-purpose untrusted outsourced cloud setting. In this example, data is encrypted using a symmetric encryption algorithm, such as Advanced Encryption Standard (AES), before being transmitted to the cloud. The encrypted data is then decrypted in the TEE and securely processed, ensuring that sensitive information remains protected throughout the computation. The model inference prediction results are then encrypted with the same key and sent back to the data provider, who decrypts them to get their final prediction.  
The TEE frameworks analyzed in this work include \cite{tee_framework8, tee_framework9, tee_framework10, tee_framework1, tee_framework2, tee_framework4, tee_framework5, tee_framework6, tee_framework7, tee_framework11, tee_framework12, tee_framework13}.

\subsection{Multi-Party Computation}
Secure Multi-Party Computation (MPC) is a cryptographic protocol that allows multiple parties to jointly compute a function while keeping their inputs private \cite{DavidEvans-2018}. The concept was pioneered by Andrew Yao in 1982 with the introduction of the Garbled Circuit Protocol, a foundational approach that allows secure function evaluation without revealing private data \cite{AndrewCYao-1982}.
Over the years, numerous MPC protocols have been developed, each tailored for different applications with varying security guarantees, computational requirements, and communication complexities. 
Some protocols prioritize complete cryptographic security under stringent assumptions, while others aim for practicality by optimizing communication and computational efficiency. 
Despite its promise, MPC comes with challenges, particularly in PPML scenarios. One of the most notable limitations is its high communication cost. Many MPC protocols require frequent exchanges of data between parties over multiple rounds of interaction, which can significantly increase latency and resource consumption, especially in settings involving large datasets or complex models. 

\begin{figure}[http]
    \begin{center}
    \includegraphics[width=0.8\linewidth]{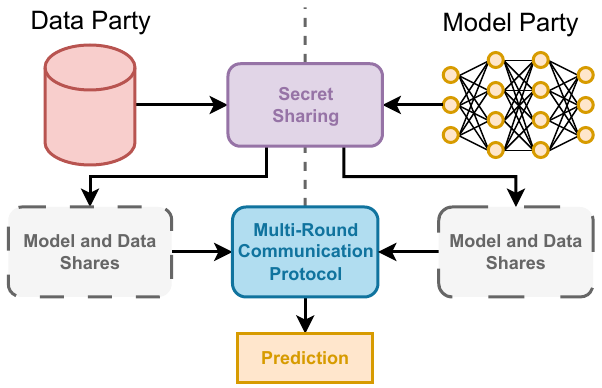}
    \captionsetup{justification=centering}
    \caption{Demonstration of a PPML inference using MPC. Two parties generate secret shares of their data and model and compute the desired functions over several communication rounds, where they exchange data and perform joint computations.}
    \label{fig:mpc_inference}
    \end{center}
\end{figure}

Figure \ref{fig:mpc_inference} illustrates an example of MPC applied to private inference or training between two parties. 
In this setup, one party holds the data while the other has the model. The process begins with both parties generating secret shares of their inputs, effectively splitting the data into parts that reveal nothing. 
These secret shares are then exchanged with the corresponding party. Once the shares are distributed, the two parties engage in secure computations over multiple rounds, using the shares to compute the desired function without exposing any private information. 
Finally, the results of the computation, whether inference outcomes or training updates, are revealed to a designated party, ensuring that the privacy of the original inputs is preserved. 
The MPC frameworks analyzed for \Prismo include \cite{JianLiu-2017, SadeghRiazi-2018, PaymanMohassel-2017, PaymanMohassel-2018, MeghaByali-2019, NishanthChandran-2019, NishantKumar-2020, PatraArpita-2020, NishatKoti-2021-Tetrad, NishatKoti-2021, TianpeiLu-2023, JunmingMa-2023}.

\subsection{Homomorphic Encryption}
Homomorphic encryption (HE) is an advanced cryptographic protocol that allows computations to be performed directly on encrypted data, ensuring sensitive information remains confidential throughout the computation process. 
This capability is especially valuable in PPML, where data privacy and model security are critical. 
Formally if given two messages \(m_1\) and \(m_2\) with an encryption function \textbf{Encrypt}, and computationally feasible functions \(f\) and \(f'\), \(f\) and \(f'\) are said to be homomorphic if:
\begin{align}
f(m_1, m_2) = f'(\textbf{Encrypt}(m_1), \textbf{Encrypt}(m_2))
\end{align}

Fully Homomorphic Encryption (FHE) \cite{gentry}  ensures that arbitrary operations on encrypted data can be performed as if operating on the original unencrypted data, preserving confidentiality throughout the computation process. 
While FHE offers high security guarantees, it introduces significant computational overhead. HE computations are resource-intensive, making them less efficient for large-scale machine learning tasks. Additionally, most FHE schemes support only linear operations, limiting their direct application to a limited subset of ML applications. 

\begin{figure}[ht]
    \begin{center}
    \captionsetup{justification=centering}
    \includegraphics[width=\linewidth]{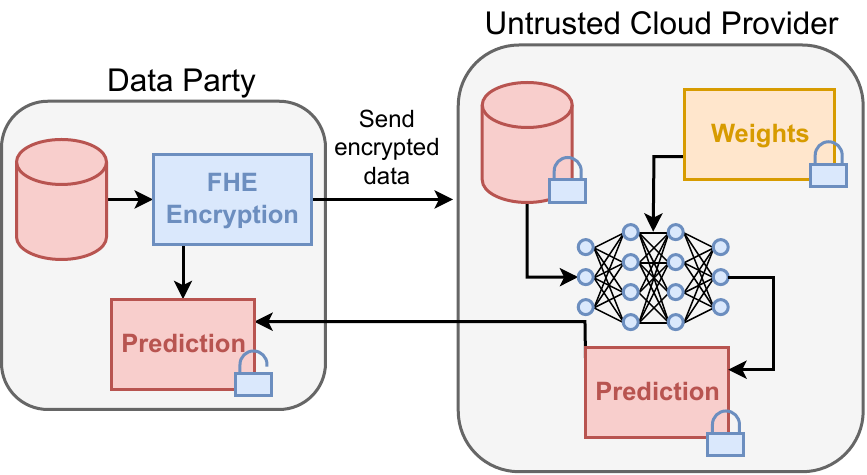}
    \caption{Demonstration of  FHE. An FHE model and encrypted weights are placed in the cloud. Clients send encrypted data, which is then inferred. Results are returned to the client, who decrypts them using their private key.}
    \label{fig:fheflow}
    \end{center}
\end{figure}

Figure \ref{fig:fheflow} illustrates how FHE can be applied in a cloud setting for PPML inference, where the model and data are kept secure. The FHE-friendly model and encrypted weights are stored on an untrusted third-party server. 
Clients send their encrypted data to the server for inference. The server directly inferences the encrypted data without decrypting it. 
After completing the inference, the server sends the encrypted results back to the client, who can decrypt them using their secret key to obtain the final output. 
This method ensures that sensitive data and the model remain private, as the server does not gain access to data at any point in the computation since everything is encrypted.
The FHE frameworks analyzed in \Prismo include \cite{crytonet, e2dm, privft, ngraph, chet, cryptodl, pyhenet, ppmlfhednn, ahec, lorenzo, ppdl, hcmain, tt-tfhe}.

\subsection{Hybrid Techniques}
Some frameworks combine multiple privacy-enhancing techniques to harness their strengths and overcome the limitations of each approach. 
Integrating the complementary capabilities of different cryptographic and privacy-preserving methods, these hybrid frameworks can provide more secure and efficient solutions for privacy-sensitive applications. 
For instance, the HT2ML framework \cite{ht2ml} integrates FHE with TEEs to improve security and computational performance. 
The hybrid frameworks analyzed in this work include \cite{Shuangyi, tee_framework3, tee_framework9, pentyala, chiron, primodchain, gazelle, ohrimenko, bost14, genoppml, ppmltsa, pysyft, ryffel2018generic, ht2ml}.

\section{PPML Frameworks Classification}
\label{sec:frameworksclassification}

Security, performance, scalability, and resource efficiency are some of the primary characteristics considered when selecting PPML solutions for real-world applications. Balancing these properties remains a significant challenge within the field. 
Thus, developing a methodology that quantitatively evaluates PPML frameworks using these factors is crucial, as it will allow users to make informed decisions based on their specific use cases and select the most suitable framework for their deployment scenarios. 
To understand the similarities, differences, and to come up with an approach for a quantitative evaluation of these frameworks, a deep study is required to identify the characteristics of the existing frameworks.
Based on these characteristics, a fine criterion can be developed to convert the features into measurable quantitative data normalized across the different frameworks, features, and scenarios.

In the initial phase of this work, we analyzed seventy-four PPML frameworks to identify generic features common across all PPML solutions that can be quantified and used as a baseline for comparison. 
For every technique in the background, we listed all the frameworks we studied, and from these works, ten essential factors were identified. 
While some of these factors enable a quantitative comparison of frameworks, others offer strong connections to define relationships between frameworks. These generic factors include ML operations, threat models, training support, datasets used, data privacy, model privacy, inference time, non-linear functions in models, model accuracy, and the open-source status. 

The ML operations, datasets, and non-linear functions are key to establishing relationships between specific use cases and frameworks, but cannot be quantified. This is because there is no measurable relationship between the different possible values of these features. 
ML operations include the convolution and fully connected layers for deep networks, methodologies such as k-means clustering and decision trees in simpler works. 
Datasets include the MNIST, FashionMNIST, CIFAR-10, ImageNet, and Private Datasets used to validate the frameworks. Non-linear functions include: ReLU, the Square function, and other low-degree polynomials. 
On the other hand,  the threat model, training and inference support, data and model privacy, and open-source status not only establish relationships but also provide a foundation for quantitative comparison of frameworks. 
Specifically, the open-source status, training and inference support, and data and model privacy can be transformed into binary features indicating whether a framework supports them or not.
We say a framework offers data privacy when it provides data protection during training or inference. In the same light, we say a framework offers model privacy when it ensures that the ML model is kept private during these stages. The threat model, model accuracy, and inference time of every framework are derived directly from the publications associated with it.

Each technique has additional features that provide valuable insights into the capabilities of it's frameworks. For FL, these features include the number of clients supported, the number of rounds needed, hardware acceleration, the types of FL methodologies supported (whether centralized, decentralized, or both), and the aggregation algorithms(e.g., federated averaging). 
In TEEs, important features include the hardware used (e.g., Intel SGX), provision of protection against attacks, such as side-channel attacks, and the presence of integrity checks. In multi-party computation, the schemes employed (such as garbled circuits) and the number of participants supported are crucial features. The protocols used by the framework are a key consideration for differential privacy. 
Furthermore, in HE, key factors include the scheme used (e.g., CKKS), the library used (e.g., OpenFHE), support for optimization techniques such as batch normalization, bootstrapping, and support for hardware acceleration.
Finally, for hybrid frameworks, we extracted the techniques used, the number of parties involved, and the hardware acceleration status of the frameworks.  

Table \ref{tab:ppmlframeworks} summarizes the number of frameworks analyzed for each technique, while Table \ref{tab:techniquefeatures} shows all the additional features extracted for every technique.
Table \ref{tab:ppmlframeworks_details} summarizes the number of frameworks exhibiting the different features discussed above. It outlines the count of frameworks for binary features, the count of frameworks categorized under the different threat models defined by \Prismo, the number of frameworks supporting different non-linear functions, and the number of frameworks capable of handling scenarios involving more than two parties. These tables do not include information on datasets, accuracies, and inference time since they can vary greatly between frameworks.

Understanding the impact of extracted features across all PPML techniques and determining which features are relevant for specific approaches is not a trivial task. It requires a thorough analysis of existing works to establish both valuable and available information. The criteria established in this analysis lay a foundation for analyzing other related works, to extract relevant information, thereby supporting a comprehensive comparative analysis and informed recommendations of all PPML frameworks.

\begin{table}[http]
    \small
    \centering
    \vspace{0.2in}
    \caption{Number of Frameworks Analyzed per Technique}
    \label{tab:ppmlframeworks}
    \begin{tabularx}{\linewidth}{|X|c|}
        \hline
        \textbf{Technique} & \textbf{Number of Frameworks} \\
        \hline
        Federated Learning & 12 \\
        \hline
        Multi-Party Computation & 13 \\
        \hline
        Differential Privacy & 9 \\
        \hline
        Homomorphic Encryption & 14 \\
        \hline
        Trusted Execution Environment & 12 \\
        \hline
        Hybrid Frameworks & 14 \\
        \hline
    \end{tabularx}
\end{table}

\begin{table}[http]
    \small
    \centering
    \caption{Additional features extracted per technique. For DP, only the Scheme was extracted as an additional feature. Number of (No.), Aggregation  Algorithms (Aggr. Algo.), Hardware Acceleration (Acc.), Integrity (Int.), Protected (Prot.)}
    \label{tab:techniquefeatures}
    \begin{tabularx}{\linewidth}{|X|X|X|X|X|}
        \hline
        \textbf{FL} & \textbf{TEEs} & \textbf{MPC} & \textbf{HE} & \textbf{Hybrid} \\
        \hline
        No. clients & Hardware & Schemes & Scheme & Techniques \\
        \hline
        No. Rounds & Prot. Attacks & No. Participants & Normaliz-ation & No. Parties \\
        \hline
        Acc. & Acc. & & Acc. & Acc. \\
        \hline
        Library & Int. Check & & Library  & \\
        \hline
         Methodo-logies  & Edge Support & & Bootstrap-ping & \\
        \hline
         Aggr. Algo. &  & &  & \\
        \hline
    \end{tabularx}
\end{table}

\begin{table}[ht]
\footnotesize
\centering
\renewcommand{\arraystretch}{1.0}
\caption{Number of frameworks exhibiting certain extracted features.}
\label{tab:ppmlframeworks_details}
\begin{tabular}{|p{0.55\columnwidth}|c|}
\hline
\textbf{Feature} & \textbf{Number of Frameworks} \\
\hline
\multicolumn{2}{|c|}{\textbf{General Features}} \\
\hline
Open Source & 32 \\
\hline
Training Support & 44 \\
\hline
\multicolumn{2}{|c|}{\textbf{Privacy Features}} \\
\hline
Data Privacy & 64 \\
\hline
Model Privacy & 47 \\
\hline
\multicolumn{2}{|c|}{\textbf{Threat Models}} \\
\hline
Malicious Adversaries & 13 \\
\hline
Semi-Honest Adversaries & 53 \\
\hline
Trusted Third-Party & 13 \\
\hline
\multicolumn{2}{|c|}{\textbf{Advanced Features}} \\
\hline
Use Non-Standard Non-Linear Functions & 20 \\
\hline
Supports More Than Two Parties & 23 \\
\hline
\end{tabular}
\end{table}

\section{\textsc{Prismo} Recommendation System}
\label{sec:framework}

Given the information extracted from frameworks, an analytical tool that enables users to comprehensively represent their specific use cases is required. Such a tool must support systematic comparison, accommodate heterogeneous privacy requirements, and remain adaptable to evolving frameworks. 
The solution must offer the flexibility, extensibility, and usability required for real-world usage. Based on these requirements, we designed  \Prismo, an extensible platform for selecting effective PPML instances. 
It approaches the selection problem by formulating it as a customizable multi-objective linear integer programming problem on different frameworks. To enhance accessibility, \Prismo features a user-friendly web-based interface, making it accessible. 
The platform offers five core functionalities: Search, Filtering, Ranking, detailed framework pages, and the ability to add new frameworks. These features ensure that users can easily represent their deployment scenarios, discover and evaluate frameworks, as well as customize and expand the range of recommended frameworks to match their needs. \Prismo is publicly available for use at \url{https://prismo.ascslab-tools.org}.

\subsection{Search} 

\Prismo search functionality is designed to balance comprehensiveness with user-friendliness, offering a high degree of customization through six key attributes. Each attribute is optional, allowing users to tailor searches to their specific use-case requirements. The attributes help map and analyze the user's requirements with the data extracted from the frameworks present in the database.
This feature enables \Prismo to map frameworks into a space that meets the users' needs. 
The six attributes used for search include PPML Technique, ML Models, Threat Model, Datasets, Training Status, and Open-Source Status. 

Users can narrow their search to frameworks utilizing specific methodologies by selecting a particular technique discussed in Section \ref{sec:background}, thus enabling users to quickly identify frameworks that implement privacy-preserving mechanisms most relevant to their needs. \Prismo also allows users to choose from its database's entire range of ML models. Users can select multiple models simultaneously, ensuring the search results encompass all frameworks relevant to their use case. This feature supports diverse use cases, from simple classifiers to advanced networks. 
Datasets are also a critical component of PPML workflows. Users can search for frameworks that use specific datasets, ensuring compatibility with their data requirements and enhancing the applicability of the identified frameworks. The \Prismo datasets are derived from the complete list of datasets used by the frameworks analyzed in this work and available in the database.

Security and Privacy considerations are paramount to \Prismo. In \Prismo, we categorize all frameworks under the three standard but distinct threat models. The first is malicious adversaries, who assume competent adversaries with unrestricted malicious intent. The second is semi-honest adversaries, focusing on adversaries who follow protocols but attempt to infer private information. The third is semi-honest with trusted third parties, which introduces a trusted intermediary to enhance security while retaining semi-honest behavior. This classification ensures comprehensive protection guarantees, allowing users to select frameworks that align with their security objectives. 
Furthermore, privacy requirements can differ across the different stages of the ML lifecycle, so allowing users to search for frameworks that meet their training state is also important. \Prismo further supports search refinement based on privacy-preserving mechanisms applied during inference only, training only, and both phases. 
Finally, \Prismo offers an option to search for frameworks based on their open-source status, enabling users to prioritize accessibility and collaboration or focus on proprietary solutions per their preferences.
By leveraging these customizable attributes, \Prismo empowers users to conduct highly targeted searches that yield relevant and actionable results tailored to their unique requirements.

\subsection{Filtering}
Filtering allows \Prismo users to further refine search results from the initial query. This feature is more useful for advanced users in tailoring results to individual needs. While filters such as hardware acceleration require no prior knowledge of PPML, more advanced filters, such as protocols and libraries, are better suited for users with a foundational understanding of the PPML technique under consideration.

    $\bullet$ \textbf{Hardware Acceleration:}  Computational cost is a well-known challenge in ML. When ML combines with intense computational techniques like HE, hardware acceleration becomes critical to enhance the speed and efficiency of computations. Currently, three primary types of hardware are widely employed in accelerating computations: GPUs, FPGAs, and ASICs. 
    These options provide varying degrees of flexibility, performance, and efficiency in accelerating PPML applications. PPML users who intend to accelerate their work can use this filter to help them target frameworks with even better-fitting hardware recommendations.
    
    $\bullet$ \textbf{Protocols/Schemes:} Various PPML techniques support different protocols for application development. Users can filter results based on the protocols of their chosen method. These results align better with the specific protocol requirements of the user's application, supporting the appropriate privacy guarantees. For instance, DP, Local-DP, and ($\epsilon$, $\delta$)-DP form the foundation of many frameworks. They provide varying levels of assurance, and understanding them can allow a user to select frameworks that align more effectively with their needs.
    
    $\bullet$  \textbf{Libraries:} Different libraries are often used to develop different PPML frameworks, particularly in techniques like HE and FL. Different libraries offer different optimizations and, in some cases, varying levels of security guarantees. The choice of library often impacts the performance and behavior of the PPML frameworks built on it. For example, HE libraries such as SEAL, OpenFHE, and HEAAN provide unique cryptographic schemes and differing performances for the schemes they implement \cite{tl2benchmarks, njungle2025safety}. 
   Thus, this filter allows users with an understanding of the landscape of a privacy-preserving technique to further prioritize frameworks based on their preferred library within that specific technique.
    
    $\bullet$ \textbf{Specific Technique Features:} It introduces critical considerations for secure computation for that technique. In FL, using a centralized server and supporting edge devices can significantly influence a user's decision. Similarly, the number of parties supported by an MPC framework directly shapes its use cases and application scenarios.


\subsection{Ranking}

After searching and filtering, \Prismo\unskip’s ability to rank potential frameworks based on user-defined importance and features is the central component of the recommendation engine.
 This ranking must be both flexible and customizable to reflect the diverse deployment scenarios of PPML through user-defined inputs. This is because we must provide an approach for users to balance their user requirements, such as trade-offs between performance, robustness,  privacy guarantees, security guarantees, scalability, and accessibility.
 To address this, \Prismo formulates the selection task as a multi-objective linear integer programming (LIP) problem \cite{schrijver1998theory}, \cite{genova2011linear}.
The LIP is particularly well-suited for this problem because it offers an approach to comprehensively represent all frameworks quantitatively by encoding multiple features. 
Moreover, LIP allows us to model the recommendations with hard constraints and weighted priorities to be specified explicitly by the user, thereby enabling highly granular control and flexibility over the recommendation process. Each PPML framework can be represented as a vector of quantifiable attributes, and the LIP solver can efficiently navigate this multidimensional decision space to identify the most appropriate frameworks under the given constraints. 
Unlike heuristic or rule-based ranking methods, which may oversimplify trade-offs or lack transparency, LIP provides analytically provable optimality for selected configurations and scales well with increasing complexity and number of frameworks. 
This makes it the most robust and theoretically grounded method for ensuring that recommendations are not only aligned with user preferences but also analytically justified. Thus, LIP forms the mathematical backbone of \Prismo’s framework ranking engine, ensuring that its rankings are both interpretable and optimal with respect to user-defined criteria.

\Prismo uses two approaches for ranking frameworks: the default approach and the user-optimization approach. These approaches ensure that \Prismo recommends the best-fitting solutions for all use cases.
In these approaches, the LIP is constructed based on six primary factors as discussed below:

     $\bullet$ \textbf{Verifiable Results (Performance Evaluation):}  We assess the performance of various PPML frameworks in \Prismo by setting them up locally and running different examples primarily to reproduce the results stated in the publication. In some cases, we develop custom models to validate the accuracy of the frameworks. The accuracies derived from the frameworks serve as the performance baseline and form the recommendation system's foundation. Frameworks exhibiting superior performance are ranked highest in search results, followed by those with lower performance. Frameworks that cannot be validated due to technical setup issues are ranked the least. To standardize performance comparisons, the accuracy obtained during each framework's validation is divided by the maximum accuracy achieved for the specific dataset in the \Prismo database. Points are then assigned to each framework by normalizing the result of this operation to a value within the range $[0, 1]$, as shown in the  Equation \ref{eq:accuracy}. This method of normalizing results ensures fair and consistent ranking across different frameworks and datasets. 
    \begin{align}
        \label{eq:accuracy}
            p_i = \frac{1}{DN} \sum_{d=1}^D \sum_{j=1}^N \frac{r^{(d)}_j}{\max\limits_k (r^{(d)}_k)}.
    \end{align} 
    $p_i$ is the accuracy rank of a framework $i$, $D$ is the number of datasets used by framework $p_i$, $N$ is the total number of results from framework $i$, $r^{(d)}_i$ is the accuracy of framework $i$ on dataset $d$, $\max\limits_k (r^{(d)}_k)$ is the maximum accuracy obtained by a framework $k$.

    In addition to the accuracy metrics, which are integrated into \Prismo's ranking algorithm as the performance measure, we also record and display resource utilization information of frameworks. This information includes memory usage and communication overhead, providing users with a more comprehensive understanding of each framework's efficiency in deployment scenarios. Memory Usage is often a critical bottleneck in HE frameworks, while Communication Overhead is particularly relevant for MPC and FL frameworks. 
    Furthermore, Docker images are created from the verified frameworks with starter examples and detailed usage instructions. This information is all avaliable on the \Prismo website at \url{https://prismo.ascslab-tools.org}. These resources are valuable for users looking to quickly prototype and experiment with PPML frameworks, especially since many of the frameworks are research-based and lack comprehensive documentation, thus requiring technical expertise to set up, which is usually not widely available.
    
     It is worth noting that, while the metrics recorded from this assessment are currently given to users in a raw format, quantifying them for ranking in \Prismo is still an active research problem. This is mainly due to the diverse architectures, network configurations, and model designs employed across the different frameworks. These variations make consistent quantification and normalization of information extremely challenging. To the best of our knowledge, there are no existing methods that can assist in normalizing different ML information, such as datasets and models. Thus, a systematic approach is required to carefully weigh and calibrate this information, taking into account all the differences and chaos in the different applications. 
     
    $\bullet$ \textbf{Published Results (Robustness):} 
    The second factor influencing our ranking algorithm is the accuracy disclosed in the published papers of each framework. 
    Frameworks with higher reported inference accuracy are ranked higher, while those with lower accuracy are ranked accordingly. To assign points, we standardize and calculate scores for accuracy the same way we did with Verifiable Results in Equation \ref{eq:accuracy}. We then calculate the average of their normalized accuracies across datasets. This approach accounts for the varying datasets used across models. By incorporating published results, our ranking engine is strengthened by accounting for variations in framework availability while ensuring that frameworks with strong academic validation are highlighted. This approach provides a balance for robustness in framework recommendations.

    $\bullet$  \textbf{Model and Data Privacy (Privacy Guarantees):} The primary objective of PPML is to ensure model and data privacy during training and inference. Frameworks that provide comprehensive privacy protections for both data and models are inherently more secure than those that offer only one of these protections. This dual-layered approach addresses two critical dimensions of privacy: protecting sensitive data from exposure during training and inference and preventing leakage of proprietary model information, which can be just as valuable. 
    Frameworks that deliver data and model privacy are awarded $1$ point, while those that focus on just one (either data privacy or model privacy) receive $0.5$ points. This distinction underscores the importance of holistic privacy guarantees. By prioritizing frameworks with data and model privacy, we promote a more reliable approach to privacy in ML applications.
    
    $\bullet$ \textbf{Threat Model Protection (Security Guarantees):} The threat model protection of a framework is a critical component of its privacy and security guarantees. Different frameworks are designed to defend against various adversaries, typically distinguishing between malicious, semi-honest adversaries, and semi-honest adversaries using a trusted third party to assist in the protection. The different threat model protections are awarded points as shown in Table \ref{tab:threatranking}. The points are given to frameworks based on the strength of the security guarantees offered through their threat model. For frameworks that offer different levels of protection, the highest level is considered during the evaluation.

    \begin{table}[http]
        \centering
        \small
        \caption{Threat Model Protection Points for Framework}
        \label{tab:threatranking}
        \begin{tabularx}{\linewidth}{|X|c|}
            \hline
            \textbf{Threat Model Protection} & \textbf{Points} \\
            \hline
            Protection against Malicious Adversaries & 1.00 \\ 
            \hline
            Protection against Semi-honest Adversaries & 0.75 \\ 
            \hline
            Semi-honest Adversaries with Trusted Third-party & 0.50 \\
            \hline
        \end{tabularx}
    \end{table}

    $\bullet$ \textbf{Training and Inference Support (Scalability):} The scalability of PPML frameworks often reflects the features they support, particularly in training and inference. The ability to support training and inference requires more extensive development and infrastructure, making these frameworks more scalable and robust in handling complex tasks. Frameworks that can handle training and inference are better equipped to manage the complete ML pipeline, offering greater flexibility and security throughout the process. By supporting both stages, these frameworks ensure that data privacy and model security are maintained across the entire lifecycle of model deployment, from learning to prediction. Therefore, frameworks that support training and inference are awarded $1$ point, while those that support only inference are given $0.5$ point. 

    $\bullet$ \textbf{Open-source Status (Accessibility):} This plays a role in our search ranking as open-source frameworks are typically more accessible, offer greater flexibility, and benefit from stronger community support. If the open-source field is not specified in the search, open-source frameworks are generally ranked higher in \Prismo. Open-source frameworks are often subjected to continuous improvements and updates driven by the community, which can lead to better long-term stability and adaptability. They also enable users to inspect and modify the code, ensuring transparency, trust, and openness. As such, open-source frameworks are awarded $1$ point, while closed-source frameworks receive $0$ points.
    
For all frameworks obtained from the search, the points assigned become weights to be used in the ranking algorithm. Our methodology balances security, performance, robustness, usability, and accessibility to give users a clear indication of the best-fitting framework. Linear Integer Programming (LIP) is used in a multi-object setup to formulate the problem and provide the user with personalized recommendations. A LIP optimization problem is generally formulated as shown in equations \ref{eq:firstset} and \ref{eq:firstsec}.
\begin{align}
\label{eq:firstset}
    \max& \ c^\intercal x \\
    \label{eq:firstsec}
    \text{Subject to }& Ax \leq b.
\end{align}
where $m, n \in \mathbb{N}$, $x \in \mathbb{Z}^n$ is the solution vector, $A \in \mathbb{R}^{m \times n}$ and $b \in \mathbb{R}^m$ define the constraints, and $c \in \mathbb{R}^n$ is a vector of weights \cite{KrasimiraGenova-2011}.

We adopt the LIP setup to our setting, where $x$ is a set of features such as verifiable results, published results, open-source status, model and data privacy, and threat model protection for each framework. $c$ is a set of associated weights for each of the frameworks. The weights, $c$, are equal to the points discussed above, allowing all parameters of the frameworks to weigh equally in the ranking.
Constraints in our system work differently than in a traditional LIP as the problem definition boils down to an unconstrained maximization problem solved as shown in Equation \ref{eq:firstset} calculated for each framework.

\subsubsection{Default Approach}
This approach is applied by default to search results or filtered outcomes of \Prismo to rank the frameworks before they are sent to the frontend. 
In this approach, the value of $x$ for all frameworks is set to a median value of \(0.5\), customizable in the user-optimization approach.  Let $z_i$ be the sum of weights accumulated by framework $i$. $f$ is the recommended framework, which is the framework with the maximum score obtained from the summation of points.
As an exemplary use case to understand the algorithm, we assume we have frameworks denoted as $z_1$, $z_2$, and $z_3$. These frameworks have different threat models: semi-honest, malicious, and semi-honest, respectively. These frameworks have published accuracy of \(92\%\), \(97\%\) on the MNIST with maximum accuracy on the system of $97\%$, and \(81\%\) respectively on the CIFAR-10 dataset, which has a maximum accuracy of $88\%$ on the database. Also, $z_1$ is open-source, $z_2$ is not open-source, and $z_3$ is open-source. Additionally, $z_1$ and $z_2$ offer just inference, while $z_3$ offers both inference and training. To determine the best-fitted solution for recommendation, let us construct LIP equations for this scenario: 
\begin{align}
    z_1 &= 0.5x_1 + 0.94x_2 + 1.0x_3 + 0.5x_4 \\
    z_2 &= 1.0x_1 + 1.00x_2 + 0.0x_3 + 0.5x_4 \\
    z_3 &= 0.5x_1 + 0.92x_2 + 1.0x_3 + 1.0x_4
\end{align}
Since our default value of \(x_i\) is \(0.5\), the sum for each framework is calculated, and frameworks are ranked as below:

{\small
\begin{align}
z_1 &= 0.5\cdot0.5 + 0.94\cdot0.5 + 1.0\cdot0.5 + 0.5\cdot0.5 = 1.47 \\
z_2 &= 1.0\cdot0.5 + 1.00\cdot0.5 + 0.0\cdot0.5 + 0.5\cdot0.5 = 1.25 \\
z_3 &= 0.5\cdot0.5 + 0.92\cdot0.5 + 1.0\cdot0.5 + 1.0\cdot0.5 = 1.71
\end{align}
}

In this case, $z_3$ emerges as the recommended framework because its LIP result surpasses the other two frameworks analyzed. Despite having the lowest accuracy value, $z_3$ stands out as the only framework that supports training and is open-source. These additional features significantly enhance its practicality and accessibility. 
Furthermore, despite the strong threat model and reasonable accuracy of $z_2$, the framework is not open-source, which greatly limits its overall utility and user accessibility.  $z_1$ is an average framework compared to $z_2$ and $z_3$. It offers neither standout performance nor any unique advantages. This recommendation highlights how \Prismo balances all the features in the LIP.

\subsubsection{User-optimization Ranking Approach}
In the default approach, all the factors of $x$ contribute equally in ranking. It is used when a user provides basic information through search and filtering.  In \Prismo, the points attributed to the weights of $x$ are generalized to make the most educated guess on all frameworks. But  \Prismo gives users sophisticated control over the ranking algorithm through the customization of all parameters that weigh the LIP definition.

\textbf{Factors:} In the default approach described above, a median value of \(0.5\) is assigned as the default value to all $x_i$. This is generic, as we assume that all the factors contribute equally to the final solution. 
In some cases, this assumption is not sufficient, thus
\Prismo allows the user to calibrate the values of $x$ between $[0, b]$, where $b$ is chosen to normalize the range of weights per feature to allow fair comparisons. We set the value of $b = 1.0$ to give the user a reasonable range for calibration with equal positive and negative influence on factors. 
 Thus, the median value of \(0.5\) still works for the default value of $x_i$. The value of $x_i$ directly affects the impact that factor has on the weight of $z_i$ in the maximization problem. For easy readability at the frontend of \Prismo, the value of $b$ is scaled by $10$.

Let us add more constraints to the use case discussed in the default approach above. Let us assume that the user is more particular about security, so they change the threat model protection factor weight to \(0.8\). Additionally, the user wants to pay more attention to the robustness of frameworks; thus, they change the accuracy impact factor value to \(0.7\).   Lastly, the user is not concerned about the open-source status of the framework, thus changing its factor value to \(0.2\).
Re-calculating the sums of $z_i$ results in: 
{\small
\begin{align}
    z_1 &= 0.5*0.8 + 0.94*0.7 + 1.0*0.2 + 0.5*0.5 = 1.508 \\
    z_2 &= 1.0*0.8 + 1.00*0.7 + 0.0*0.2 + 0.5*0.5 = 1.75 \\
    z_3 &= 0.5*0.8 + 0.92*0.7 + 1.0*0.2 + 1.0*0.5 = 1.74 
\end{align}
}
In this case, $z_2$ emerges as the recommended framework ahead of $z_3$. The emphasis of the threat model impacts $z_2$ more than the other two frameworks. Further, a change in the impact of accuracy also increases its overall performance slightly better than how it affects the other two frameworks. Though the framework is not open-source and does not support training like $z_3$, its marginally better performance and threat model make it the most suitable for this case, where the user is concerned about these two factors and cares very little about the open-source status of the project. 
From this generalization, the LIP equation used by \Prismo is written as shown in Equation \ref{eq:generalip}.
\begin{align}
\label{eq:generalip}
    z =  \max \text{  } [ d^\intercal \odot \ c^\intercal ] x_i
\end{align}

where \(\odot\) represents the Hadamard product. 
The pseudocode in Algorithm \ref{alg:optimizedranking} shows how the recommendation engine of \Prismo works using this customizable equation. This high level of customization ensures that the platform meets the diverse needs of users, providing a robust tool for finding the most appropriate frameworks in PPML for diverse use cases.
%
\begin{algorithm}
\caption{Ranking Algorithm}
\label{alg:optimizedranking}
\flushleft
\textbf{Input}: Set of frameworks $\{x_i\}_{i=1}^m$. User-defined weight vector $c$. Point vector $d$. \\
\textbf{Ensure}: $x_i \in \mathbb{Z}^n$, $c \in \mathbb{R}^n$, with $m, n \in \mathbb{N}$. \\
\textbf{Output}: A list of frameworks ranked in ascending order.
\begin{algorithmic}[1]
    \STATE $rankedFrameworks \gets$ empty list
    \FOR{$j \in \{1, 2, \ldots, m\}$}
        \STATE $w \gets c$ 
        \IF{$d$ is provided}
            \STATE $w \gets d \circ c$ \COMMENT{Elementwise product}
        \ENDIF
        \STATE $rank \gets w^\top x_j$ \COMMENT{Dot product}
        \STATE Insert $(x_j, rank)$ into $rankedFrameworks$ in ascending order of $rank$
    \ENDFOR
    \STATE \textbf{return} $rankedFrameworks$
\end{algorithmic}
\end{algorithm}

\subsection{Framework Page}
After users search and filter results for a specific query, they can access a dedicated page for every framework displayed in their results. This page is a detailed resource, offering an overview of the framework. It includes essential information such as the authors' details, an abstract of the publication, links to helpful resources, as well as additional context and support. Also, it outlines critical framework specifics, such as the threat model, data and model privacy protections, training and inference capabilities, open-source status, hardware acceleration, and technical aspects like the number of parties supported for MPC frameworks. 

In addition to these details, this page summarizes the framework's published results in an easy-to-digest format. Where available, it also includes the verifiable results validated by our team with insightful comments on our findings. The results are displayed with key details such as the datasets used, the ML model implemented, training and inference accuracies, and resources used, such as run times, memory usage, and communication overhead. 
To ensure reproducibility and facilitate experimentation, we provide a Docker image of the framework with our work. This dedicated page is essential because it consolidates all relevant information into one easily accessible location, helping users make well-informed decisions about which framework best suits their needs. The page significantly enhances the user experience by offering detailed technical specifications, verified performance results, and reproducibility tools. It ensures transparency, fosters trust, and supports a seamless transition from exploration to practical application, making it an invaluable resource for PPML developers.

\subsection{Adding PPML Frameworks to \textsc{Prismo}}
In this work, we used a database of the most relevant frameworks currently available in the field of PPML. Looking ahead, we expect the continuous emergence of new frameworks, and to facilitate this growth, we have implemented a feature that allows authors to submit their work for inclusion in \Prismo. This submission process is streamlined through an intuitive and detailed form, making it accessible to all contributors. Each submission shall undergo a review process, during which the framework’s results and supporting information are reviewed. Once approved, a framework is integrated into \Prismo and added to the recommendation engine.
This feature is crucial as it ensures that all frameworks, regardless of origin, have an equal opportunity to be considered and recommended to users. This feature also plays a vital role in \Prismo\unskip's ongoing development by addressing gaps or omitted frameworks that may have been overlooked in the initial research phase. 
By enabling seamless contributions and ensuring accurate, verified integration, this approach helps keep \Prismo up to date with the latest advancements in the sub-field of PPML, fostering a more comprehensive and dynamic resource for the research community.

\section{Evaluation}
\label{sec:usecase}

\subsection{Use Cases}
We examine three use cases of \Prismo and discuss its results. 
In each use case, we demonstrate the LIP equations for the top five frameworks obtained from \Prismo. We then discuss the reasoning behind their ranking by collating the results with frameworks literature, further problem definition, and justifying \Prismo's recommendations. 
First, let us define the parameter variables of our maximization problem as Threat Model Protection ($x_1$), Model and Data Privacy ($x_2$), Published Accuracy ($x_3$), Verifiable Results ($x_4$), Open-source State ($x_5$), and  Training and Inference Support ($x_6$).

\begin{figure}[http]
    \centering
    \includegraphics[width=0.9\linewidth]{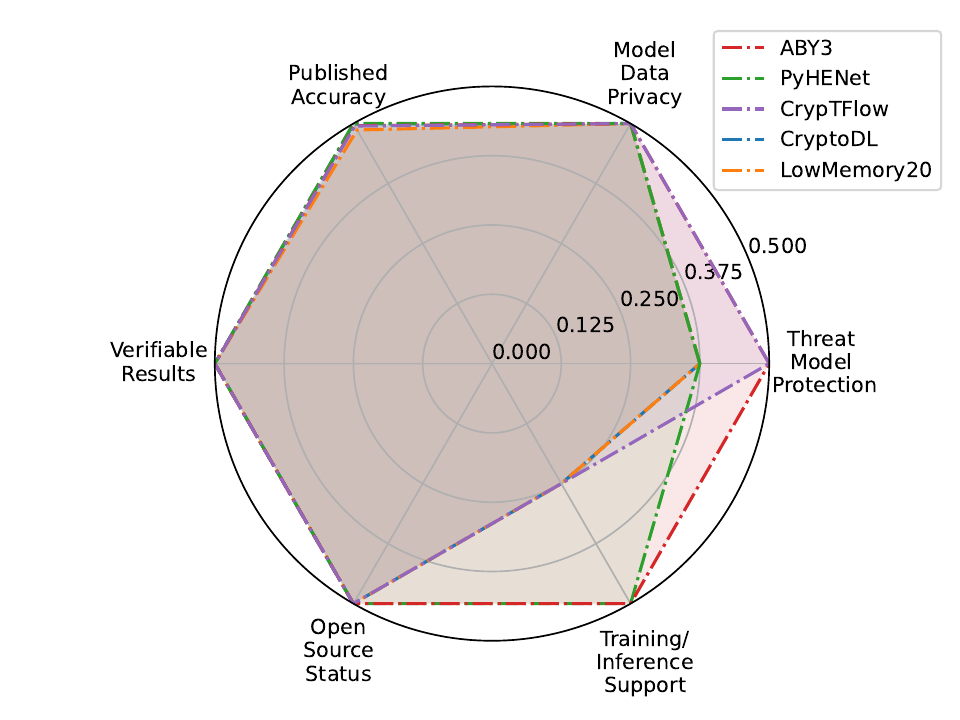}
    \caption{Radar plot representation of the top five frameworks under scenario 1}
    \label{fig:scenario1}
\end{figure}

\shortsectionBf{1st Scenario.}
We examine a use case where a \Prismo user decides to search for frameworks that support CNN inference, protect against semi-honest adversaries, and must be open-source. \Prismo's recommendation is ABY3 \cite{PaymanMohassel-2018}, with PyHENet \cite{pyhenet}, CrypTFlow \cite{NishantKumar-2020},  CryptoDL  \cite{cryptodl}, and LowMemory20 \cite{lorenzo}  making up the top five frameworks respectively.
In this case, the values of all $x_i$ equal $0.5$ as discussed in the default approach. The case generates the following equations for the top five frameworks, respectively:
{\small
\begin{align}
     & 1.0x_1 + 1.0x_2 + 0.991x_3 + 1.0x_4 + 1.0x_5 + 1.0x_6  = 2.995  \label{eq:eq1} \\
     & 0.75x_1 + 1.0x_2 + 1.0x_3 + 1.0x_4 + 1.0x_5 + 1.0x_6 =  2.875  \label{eq:eq2} \\
    & 1.0x_1 + 1.0x_2 + 0.989x_3 + 1.0x_4 + 1.0x_5 + 0.5x_6 =  2.745  \label{eq:eq3} \\
    & 0.75x_1 + 1.0x_2 + 0.98x_3 +  1.0x_4 + 1.0x_5 + 0.5x_6 = 2.624 \label{eq:eq4} \\
    & 0.75x_1 + 1.0x_2 + 0.97x_3 +  1.0x_4 + 1.0x_5 + 0.5x_6 = 2.612 \label{eq:eq5}
\end{align}
}
From observation, it is clear that the top five recommended frameworks in this case have all been verified, are all open-source, and all provide data and model privacy. The differences between them come from their threat model, published results, and support for training and inference.
ABY3 \cite{PaymanMohassel-2018} is represented by Equation \ref{eq:eq1}, and it is the recommended framework since it can be configured to protect against malicious adversaries, support both training and inference, and its published accuracy is almost $1$. Second to it is  PyHENet \cite{pyhenet}, shown by Equation \ref{eq:eq2}. It is an HE framework that provides both inference and training support as well and has the highest accuracy on the MNIST dataset presented in all PPML frameworks. Since PyHENet publication used just this dataset, it records $1$ for published results. 
It shies behind ABY3 in its threat model protection since it protects against semi-honest adversaries, offering less protection than that provided by ABY3. While  CrypTFlow represented by  \ref{eq:eq3} has a very high threat model, it only supports inference. Furthermore,  CryptoDL \cite{cryptodl} and LowMemory20 \cite{lorenzo} are very close to each other in all factors; the relative published accuracy of CryptoDL \cite{cryptodl} is slightly better than that of LowMemory20, thus giving it an edge. 
The radar plot in  Figure \ref{fig:scenario1} shows the overlaps and how the different frameworks are effectively scored for the various factors.

\begin{figure}[!]
    \centering
    \includegraphics[width=0.9\linewidth]{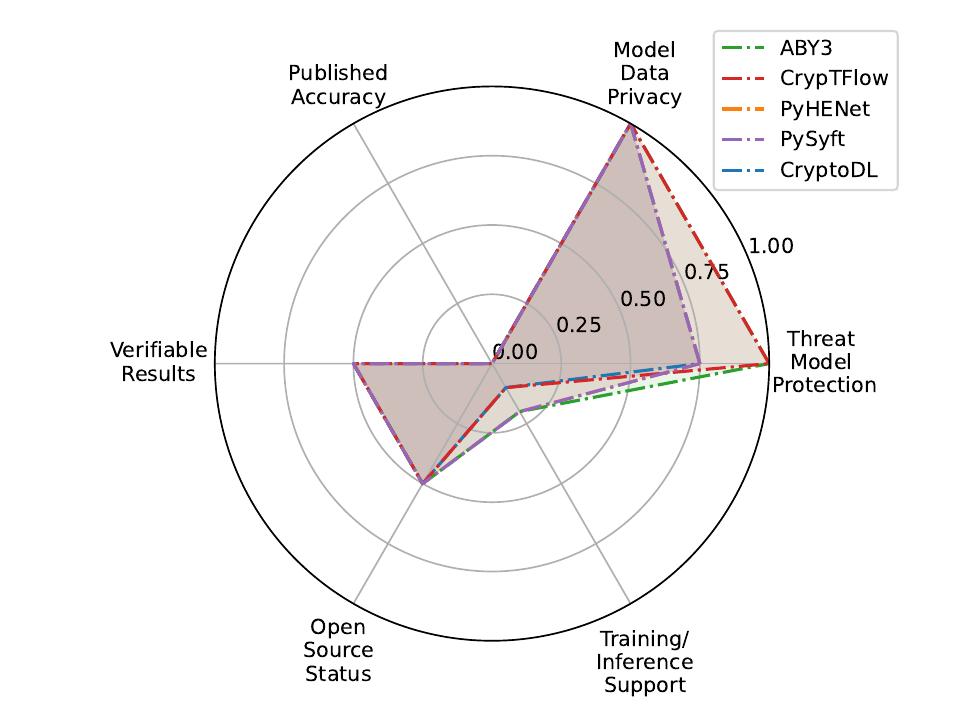}
    \caption{Radar plot representation of the top five frameworks under scenario 2}
    \label{fig:scenario2}
\end{figure}

\shortsectionBf{2nd Scenario.}
Suppose the user decides they are very interested in frameworks that offer the best all-around security and privacy and do not care about the accuracy of frameworks. 
They increase the threat model protection factor to $1$ point and the model and data privacy to $1$. They reduce the published results of the frameworks to $0$ weight. The user changes the Training and Inference Support points to $0.2$ since they are not concerned. They allow the verifiable results and open-source status at $0.5$ each as in the default case.  \Prismo's recommendation remains ABY3 \cite{PaymanMohassel-2018}, with CrypTFlow \cite{NishantKumar-2020}, PyHENet \cite{pyhenet}, PySyft \cite{pysyft}, and CryptoDL  \cite{cryptodl} making up the top five frameworks, respectively thus, demonstrating that a single framework can be the best-fitted solution for multiple PPML instances. 
The equations generated here include: 
{\small
\begin{align}
    & 1*1 + 1*1 + 0.9*0 + 1*0.5 + 1*0.5 + 1*0.2 = 3.2 \\
    & 1*1 + 1*1 + 0.9*0 + 1*0.5 + 1*0.5 + 0.5 * 0.2 = 3.1 \\
    & 0.75*1 + 1*1 + 1*0 + 1*0.5 + 1*0.5 + 1*0.2 =  2.95 \\
    & 0.75*1 + 1*1 + 0*0 + 1*0.5 + 1*0.5 + 1*0.2 =  2.95 \\
    & 0.75*1 + 1*1 + .9*0 + 1*0.5 + 1*0.5 + .5*0.2 =  2.8
\end{align}
}
From observations, it is clear that the top five frameworks in this scenario have been verified. The published results do not affect the ranking. Conversely, threat model protection and Model and Data Privacy are the factors with maximum impact.  While all frameworks protect and provide Model and Data Privacy, they process different threat models. Thus, ABY3 is retained as the recommended framework, followed by CrypTFlow.  The top two frameworks are the only ones in the results that protect against malicious adversaries. Though Training and Inference Support does not significantly impact the overall results, it gives ABY3 a slight edge over CrypTFlow since the latter only supports inference. Noticeably, PySyft \cite{pysyft} appears in the fourth place. PySyft is an FL framework with no accuracy discussed in its original publication. It also supports both training and inference and has been verified. Though the last two factors have little impact on the ranking, they place PySyft above CryptoDL.
Figure \ref{fig:scenario2} shows a radar plot of how the top frameworks behave and overlap in this scenario.

\begin{figure}[http]
    \centering
    \includegraphics[width=0.9\linewidth]{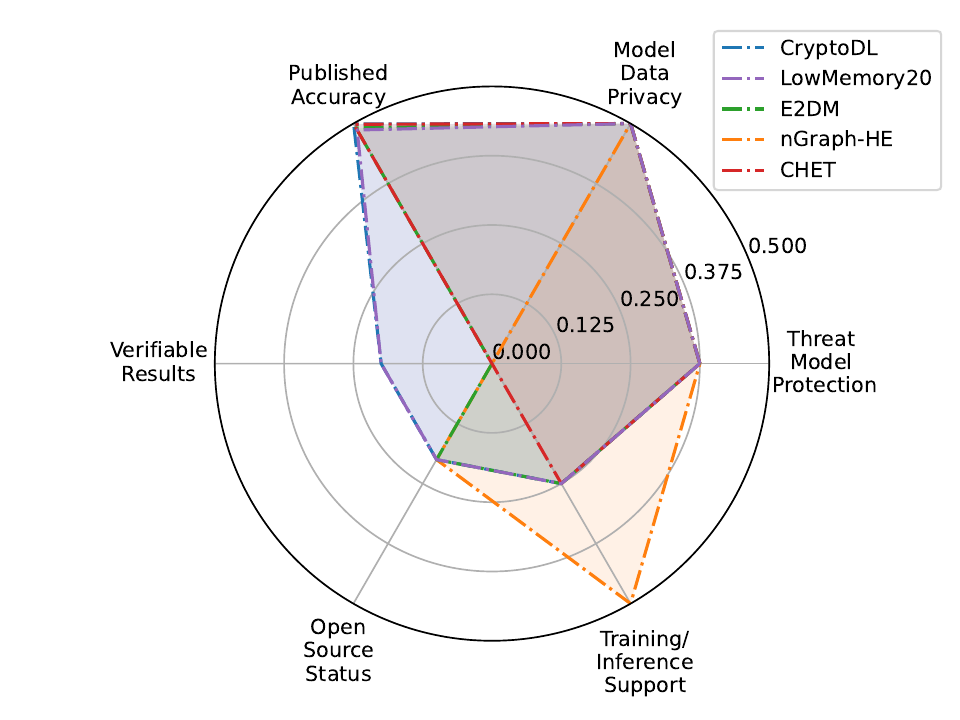}
    \caption{Radar plot showing the behavior of the top five frameworks under scenario 3}
    \label{fig:scenario03}
\end{figure}

\shortsectionBf{3rd Scenario.}
In the third scenario, we assume the user decides to use an HE Framework. The user does not care about the open-source state and changes the factor points to $0.2$. Also, they care if the results have been verified and change the factor to $0.8$. The user wants to evaluate a circuit not bounded by the set security parameters, thus filtering only the frameworks that support bootstrapping. The recommended framework is CryptoDL \cite{cryptodl} followed by LowMemory20 \cite{lorenzo}, PrivFT \cite{privft},  E2DM \cite{e2dm}, and CHET \cite{chet} as the top five frameworks. The equations generated here include: 
{\footnotesize
\begin{align}
    & 0.75*.5 + 1*.5 + 0.99*.5 + 1*.8 + 1*.2 + .5*.5 = 2.62 \\
    & 0.75*.5 + 1*.5 + 0.97*.5 + 1*.8 + 1*.2 + .5*.5 = 2.61 \\
    & 0.75*.5 + 1*.5 + 1*.5 + 0*.8 + 0*.2 + 1*.5 = 1.875 \\
   & 0.75*.5 + 1*.5 + 0.98*.5 + 0*.8 + 0*.2 + .5*.5 =  1.82 \\
   & 0.75*.5 + 1*.5 + 0.99*.5 + 0*.8 + 0*.2 + .5*.5 =  1.62
\end{align}
}
In this use case, it is clear that all the frameworks have the same threat model and offer both model and data privacy; thus, these factors do not influence the ranking. Since the open-source status of the framework is set to 0, the factor also does not affect the results. CryptoDL becomes the recommended framework above LowMemory20 simply because the framework's relatively published results are slightly better. PrivFT, E2DM, and CHET fall behind the top two frameworks mainly because they are not verified, significantly influencing the ranking scores. While PrivFT shows the best relative published results on the dataset it was trained on, it also outperforms the other frameworks since it supports training and inference.  
Figure \ref{fig:scenario03} shows a radar plot of the top five frameworks in this scenario and how they behave against the other factors.

The recommendations of \Prismo in these use cases, supported by data from relevant papers, demonstrate that \Prismo effectively fulfills its purpose. It consistently recommends the most suitable frameworks in all scenarios. In every case, we manually solved the LIP and arrived at the same recommended frameworks, which show clear advantages over alternatives, as evidenced by their performance and supporting literature. 
Table \ref{tab:scenarios} shows the points for all the frameworks discussed in these scenarios. 

\begin{table*}[ht]
    \small
    \centering
    \caption{Framework points for the six different factors}
    \label{tab:scenarios}
    \begin{tabularx}{\linewidth}{|X|X|X|X|X|X|X|}
        \hline
        \textbf{Framework} & \textbf{Threat Model Protection} & \textbf{ Model and Data Privacy} & \textbf{Relative Published Accuracy} & \textbf{Verifiable Results} & \textbf{Open-source Status} & \textbf{Training and Inference Support} \\
        \hline
        ABY3 \cite{MeghaByali-2019} & 1.0 & 1.0 & 0.991 & 1.0 & 1.0 & 1.0 \\
        \hline
        PyHENet \cite{pyhenet} &  0.75 & 1.0 & 1.0 & 1.0 & 1.0 & 1.0 \\
        \hline
        PySyft \cite{pysyft} & 0.75 & 1.0 & 0.00 & 1.0 & 1.0 & 1.0 \\
        \hline
         CrypTFlow \cite{NishantKumar-2020} & 1.0 & 1.0 &  0.989 & 1.0 & 1.0 & 0.5 \\
        \hline
        LowMemory20 \cite{lorenzo} & 0.75 & 1.0 & 0.973 & 1.0 & 1.0 & 0.5 \\
        \hline
        CryptoDl \cite{cryptodl} & 0.75 & 1.0 & 0.998 & 1.0 & 1.0 & 0.5 \\
        \hline
         PrivFT \cite{privft} & 0.75 & 1.0 & 1.0 & 0.0 & 0.0 & 1.0 \\
        \hline
         E2DM \cite{e2dm} & 0.75 & 1.0 & 0.982 & 0.0 & 0.0 & 0.5 \\
         \hline
         CHET \cite{chet} & 0.75 & 1.0 &  0.996 & 0 & 0 & 0.5 \\
        \hline
    \end{tabularx}
\end{table*}

\subsection{Prismo's Users }
\Prismo serves a broad range of stakeholders  within the PPML ecosystem, primarily falling into three groups:

\textbf{Developers and Engineers.}
Developers integrating PPML capabilities into real-world applications require practical tools to identify frameworks that align with specific operational constraints. \Prismo addresses this need by providing actionable recommendations that consider most relevant factors of PPML frameworks. The platform’s structured ranking system enables engineers to balance robustness, performance, and deployment feasibility, ultimately accelerating the development of secure and privacy-conscious ML pipelines.

\textbf{Organizations and Decision-Makers.}
Organizations cam use \Prismo to inform strategic adoption of PPML solutions that align with business goals, compliance requirements, and resource constraints. For decision-makers, \Prismo provides an evidence-driven mechanism to evaluate candidate frameworks. By distilling complex technical metrics into clear, comparable scores, the platform supports informed investment and procurement decisions.

\textbf{Researchers.}
Academic and industrial researchers will constitute a core segment of Prismo’s user base. They can use \Prismo to systematically explore the landscape of PPML frameworks, identify state-of-the-art solutions, and evaluate trade-offs. By leveraging \Prismo’s fine-grained search, filtering, and ranking capabilities, researchers can rapidly narrow down suitable frameworks for experimental studies, reproducibility efforts, and comparative analysis, thereby reducing the overhead traditionally associated with PPML frameworks related works search an setup.

Collectively, these user groups shall benefit from \Prismo’s ability to transform a fragmented and rapidly evolving PPML landscape into an accessible, structured decision-support platform. \Prismo ensures that both technical and non-technical stakeholders can select frameworks that are not only theoretically optimal but also operationally viable for their different application scenarios.

\section{Related Works}
\label{sec:related_works}
The authors of \cite{fhe_recomm} introduced an FHE recommender system for ML, employing the TOPSIS method to recommend the most suitable FHE library based on specific FHE parameters. Similarly, \cite{Yang2020} proposed a recommender system for FL, assisting users in selecting the most appropriate FL approaches and algorithms for their specific deployment needs. While these two recommendation systems focus on assisting in developing PPML tools using the HE and FL techniques, \Prismo has a much broader scope, and it focuses on real-world PPML deployment instances and scenarios.

Additionally, numerous surveys and Systematization of Knowledge (SoK) works have delved into PPML as its significance continues to rise in industry and academia. For instance, \cite{Tanuwidjaja} offered a thorough survey on deep learning techniques that facilitate access to sensitive data, providing an in-depth evaluation of various methods along with their respective strengths and weaknesses. Similarly, \cite{zohra} explored various PPML techniques, highlighting the libraries associated with each process and demonstrating how these libraries can be leveraged to develop applications that meet specific security requirements. Other works, such as \cite{BOULEMTAFES202021}, \cite{Rubaie}, and \cite{runhua}, focus on the challenges, threats, and potential solutions to enhance PPML, shedding light on the complexities of implementing privacy-preserving solutions in ML systems. While these general surveys offer valuable insights into the various PPML techniques, they primarily provide broad overviews of the methods without delving into the specific frameworks that implement them. 
PPML techniques have also been reviewed in studies such as \cite{9936637}, which analyzes frameworks for deep learning with FHE. \cite{zhang} reviewed research based on PPML frameworks. \cite{ZHANG2021106775} provided a comprehensive survey on FL, \cite{Baraheem} surveyed DP applications in ML, and \cite{Mo_2024} published an SoK on confidential computing with ML, focusing on PPML using TEEs. While these studies cover most frameworks in the different PPML techniques, their classifications are also limited, often grouping works under just one or two specific classes. Furthermore, they focus on a narrow set of features, and their analysis also remains confined to individual techniques, thus cannot be used in comparing frameworks across different PPML techniques. 

While PPML frameworks aim to deliver solutions for specific PPML instances, PPML compilers focus on simplifying the development process by offering a high-level abstraction for techniques and even automating code generation in some cases \cite{9218508, CostCO}.  In contrast to both frameworks and compilers, \Prismo is a recommendation system that guides users in selecting the best-fitting PPML solutions for specific application instances. By bridging the gap between diverse PPML tools and real-world needs, \Prismo empowers users to efficiently identify and apply the most suitable frameworks and techniques in their applications and use cases.

Unlike the prior works, our research evaluates seventy-four PPML frameworks across different techniques and identifies ten common factors that can be extracted from all PPML works for comparison. These factors provide a thorough foundation for PPML framework analysis, facilitating quantitative and cross-technique comparisons. To the best of our knowledge, we introduce the first recommendation system tailored to all PPML frameworks. \Prismo helps users analyze frameworks, rank them, and offer a comprehensive overview of all frameworks, allowing users to grasp their key features quickly. Furthermore, we present various open-source repositories containing Dockerized, tested, and well-documented versions of several PPML frameworks, offering a practical and guided resource for developers and researchers in the field of PPML.

\section{Discussions and Conclusion }
\label{sec:conclusion}

In this work, we proposed a recommendation system for PPML frameworks called \Prismo. We analyze seventy-four PPML frameworks and establish a list of ten common factors. We then use these factors to establish fine-grained criteria for extracting information from all PPML frameworks for both comparative and quantitative comparisons across all techniques. The information extracted from the seventy-four frameworks provides a basis for the development and evaluation of \Prismo for real-world adoption.   
Additionally, we provide open-source repositories that include starter examples and setup instructions for various PPML frameworks. 
This resource is a foundational tool, enabling practitioners to explore and adopt the latest advancements in PPML. 
 \Prismo is made available for use as a web application through its user-friendly interface accessible at \url{https://prismo.ascslab-tools.org}.

Although \Prismo provides a comprehensive recommendation tool for various PPML use cases, its ranking algorithm is currently constrained to the quantifiable data collected in this study. 
Looking forward, our research efforts for \Prismo include integrating a robust review mechanism for users to augment our recommendation engine's efficiency, further enhancing the decision-making processes.
\Prismo is a significant stride toward enhancing the landscape of PPML by fostering accessibility, innovation, and user empowerment. We aim to catalyze advancements in the field, driving toward a future where privacy and ML seamlessly converge.

\bibliographystyle{IEEEtran}
\bibliography{paper}

\appendix
\section{Appendix}

\subsection{\textsc{Prismo} Implementation}
\Prismo is a highly customizable platform designed with scalability in mind. It has robust features such as search and filtering, the ability to add new frameworks, dedicated inner pages, and starter kits that empower users to explore and engage with PPML technologies easily. The platform provides full autonomy over the information within the system through a user-friendly interface, ensuring a seamless experience for developers and researchers.

\begin{figure}[http]
    \begin{center}
    \includegraphics[width=\linewidth]{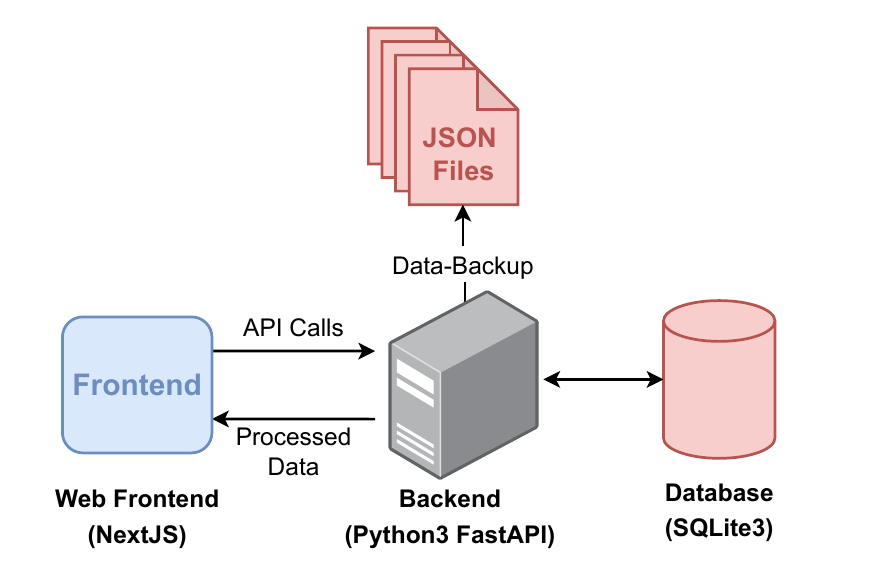}
    \caption{\Prismo System Architecture}
    \label{fig:guardianmlarch}
    \end{center}
\end{figure}

\Prismo architecture is built around three core components: the database, backend API, and frontend, as illustrated in Figure \ref{fig:guardianmlarch}. Information entered through the frontend is filtered and validated before being submitted to the backend for processing. Data is then retrieved from the database, processed, and returned to the frontend for display. This streamlined process ensures that users can easily access and interact with the necessary information. In addition to its core features, \Prismo includes an essential data backup mechanism for added protection. A corresponding JSON file is generated and stored in an external system for every framework entered into the database. These backup files are a safeguard and can be used anytime to restore the system to its previous state. This backup feature ensures the system's integrity and provides an extra layer of security against potential data loss, further enhancing the platform’s reliability and robustness.

\begin{figure}[http]
    \begin{center}
    \includegraphics[width=\linewidth]{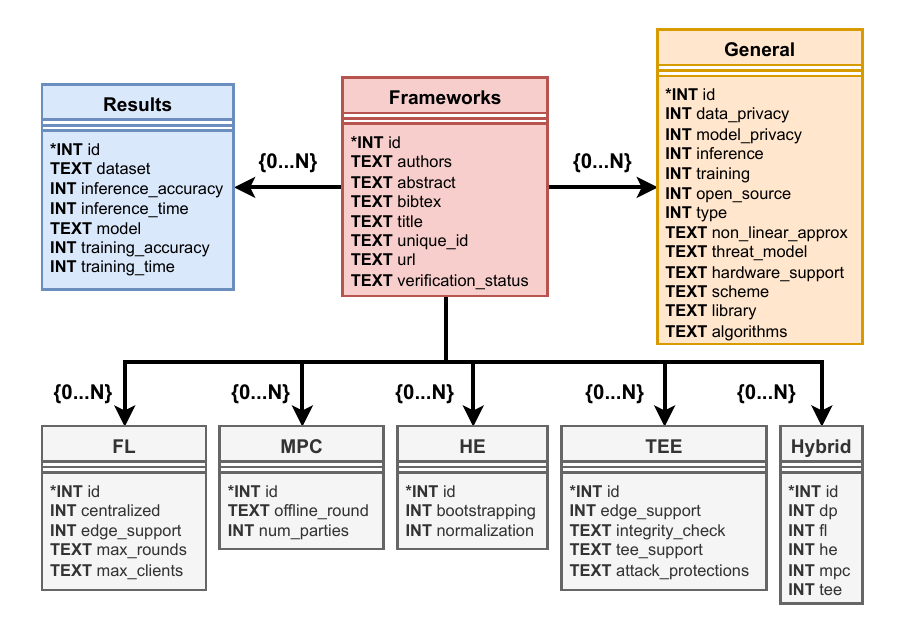}
    \caption{Simplified ER Diagram of the Database}
    \label{fig:er_diagram}
    \end{center}
\end{figure}

We use SQLite3 for the database due to its lightweight nature, modern SQL capabilities, and exceptional efficiency in managing data. As illustrated in Figure \ref{fig:er_diagram}, the Entity-Relationship (ER) diagram provides the database structure. Due to its simplicity and performance, SQLite3 is particularly suited for \Prismo, making it an ideal choice for a scalable platform. 

We have chosen Python’s FastAPI to develop the API for the backend, leveraging its powerful features that streamline the creation of Python-based endpoints. FastAPI also integrates seamlessly with SwaggerHub, providing automatic, detailed API documentation. This integration simplifies API management and ensures rapid deployment of new features, which aligns with \Prismo's focus on scalability, developer productivity, and maintaining a seamless user experience. 
The frontend is built using Next.js, simplifying the development of complex user interfaces while maintaining a clear separation between the application's state and its views. This ensures flexibility and maintainability. Additionally, Next.js supports server-side rendering, improving page load times and overall performance. This feature is critical for providing a smooth, responsive user experience, especially when handling large datasets or dynamic content.

\newpage
\subsection{Symtem Design and Live Visuals}

\begin{figure}[http]
    \centering
    \includegraphics[width=1.0\linewidth]{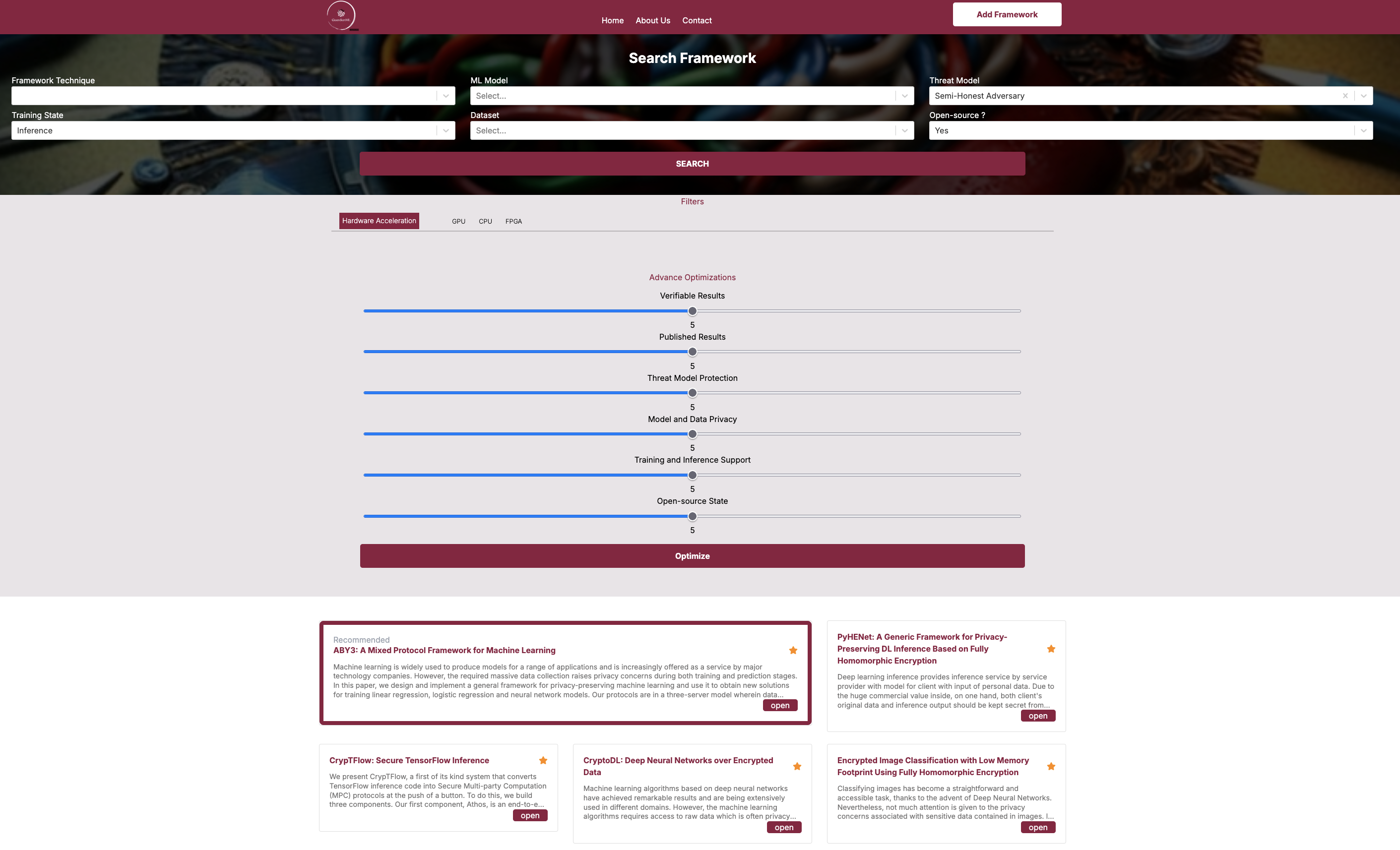}
    \caption{\Prismo Results from Scenario 1}
    \label{fig:sc1}
\end{figure}

\begin{figure}[http]
    \centering
    \includegraphics[width=1.0\linewidth]{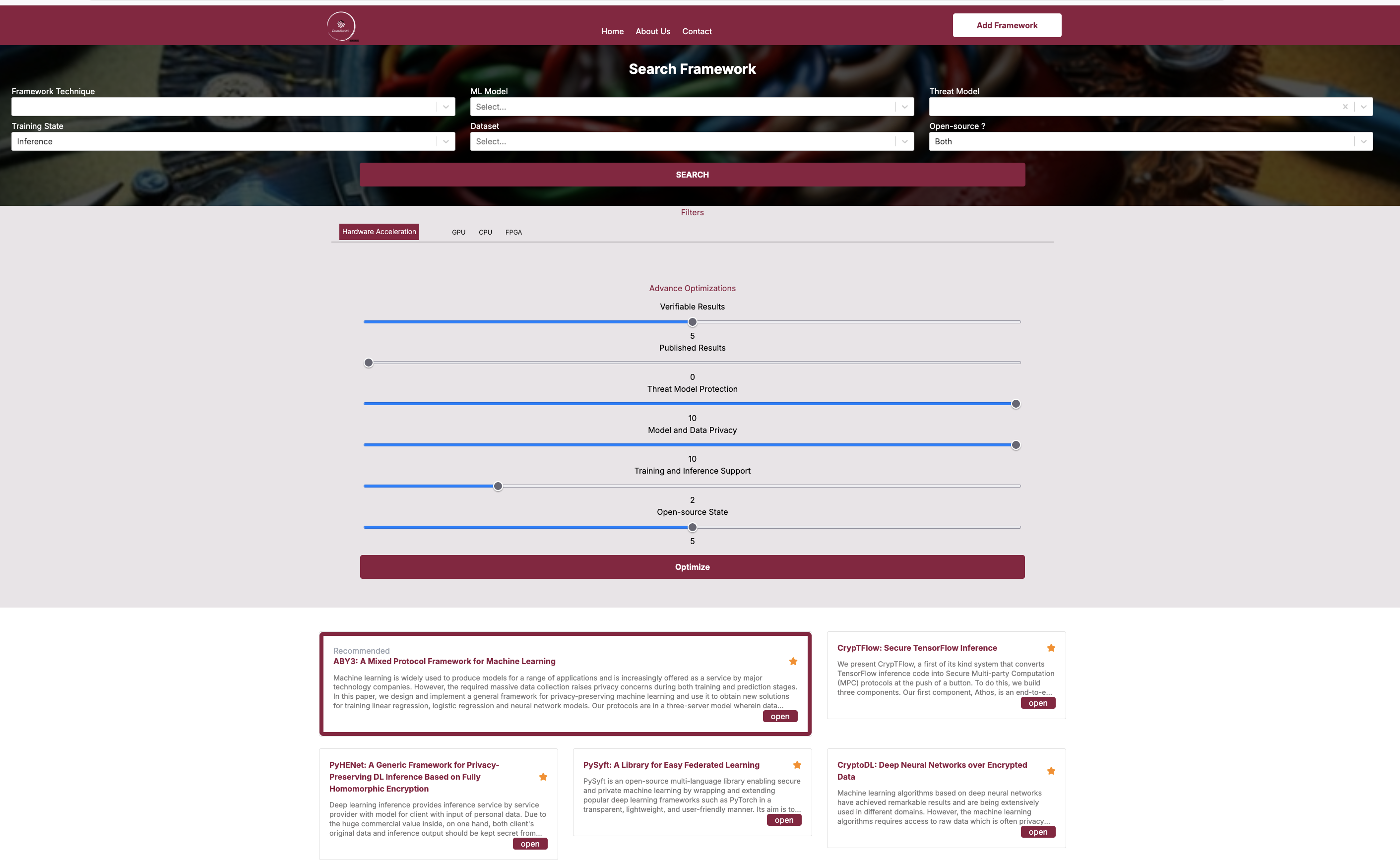}
    \caption{\Prismo Results from Scenario 2}
    \label{app:sc2}
\end{figure}

\begin{figure}[http]
    \centering
    \includegraphics[width=1.0\linewidth]{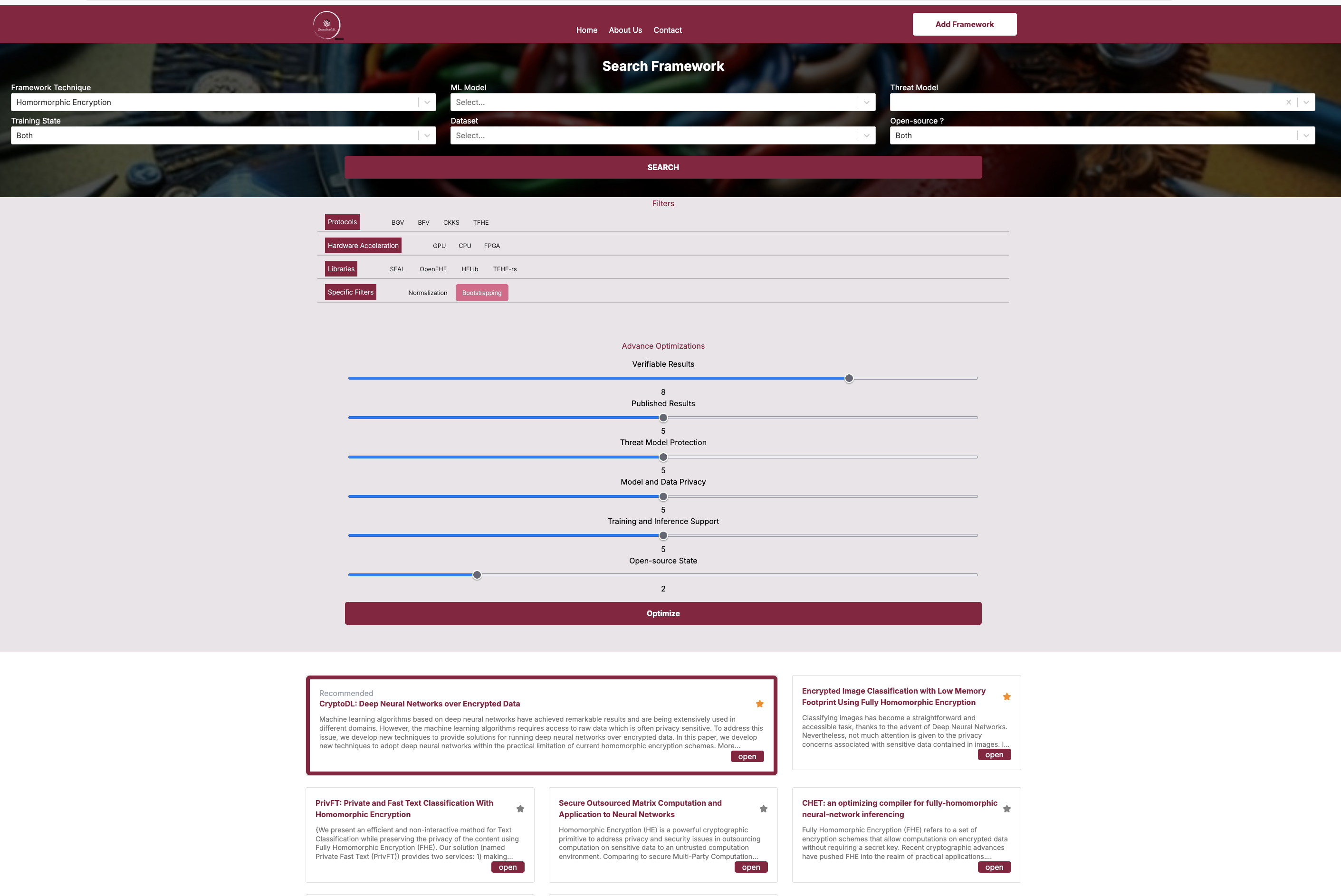}
    \caption{\Prismo Results from Scenario 3}
    \label{app:sc3}
\end{figure}

\begin{figure}[http]
\vspace{0.4in}
    \centering
    \includegraphics[width=1.0\linewidth]{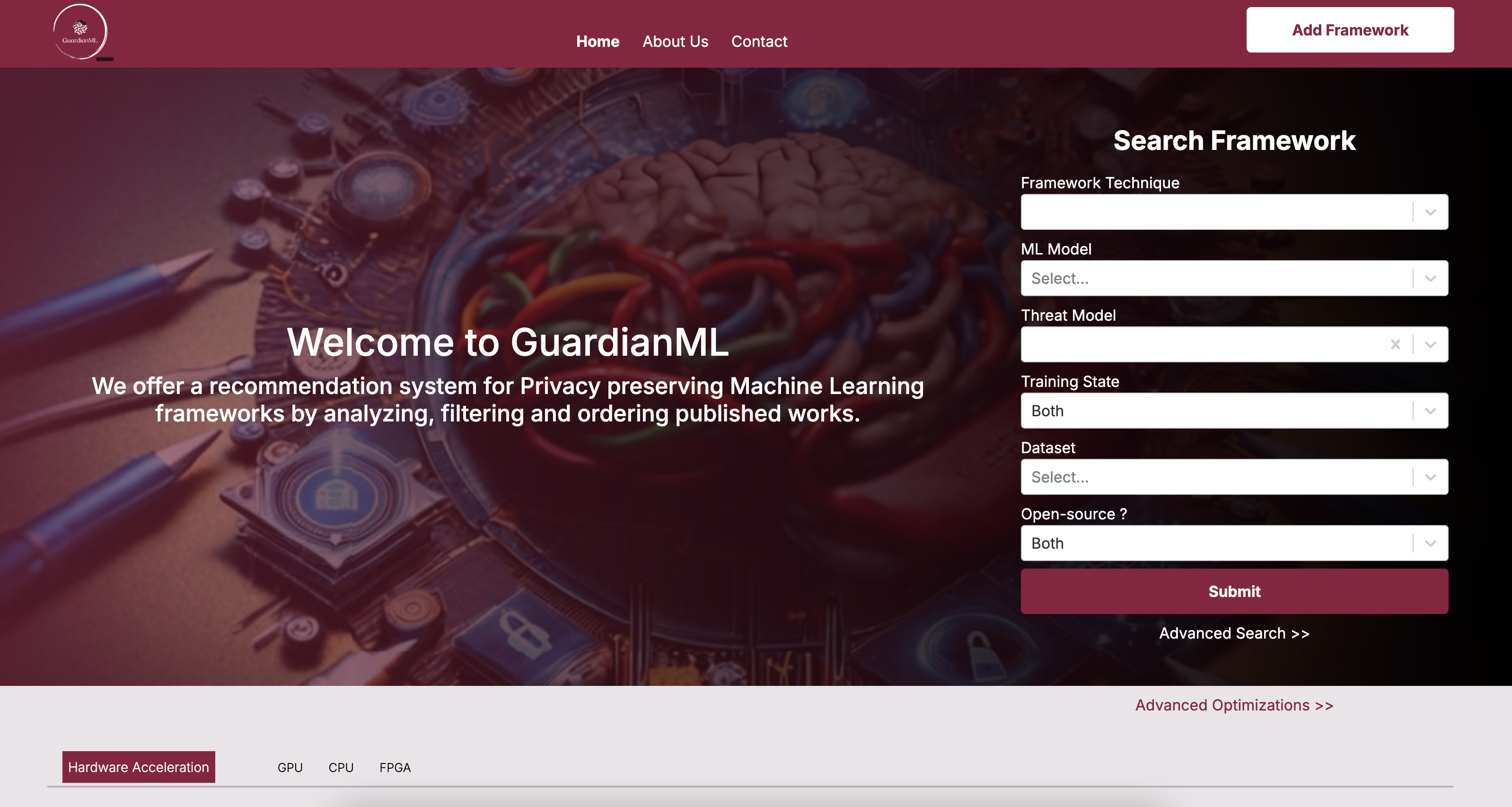}
    \caption{\Prismo Welcome page}
    \label{fig:appendixwelcome}
\end{figure}

\begin{figure}[http]
    \vspace{0.2in}
    \centering
    \includegraphics[width=1.0\linewidth]{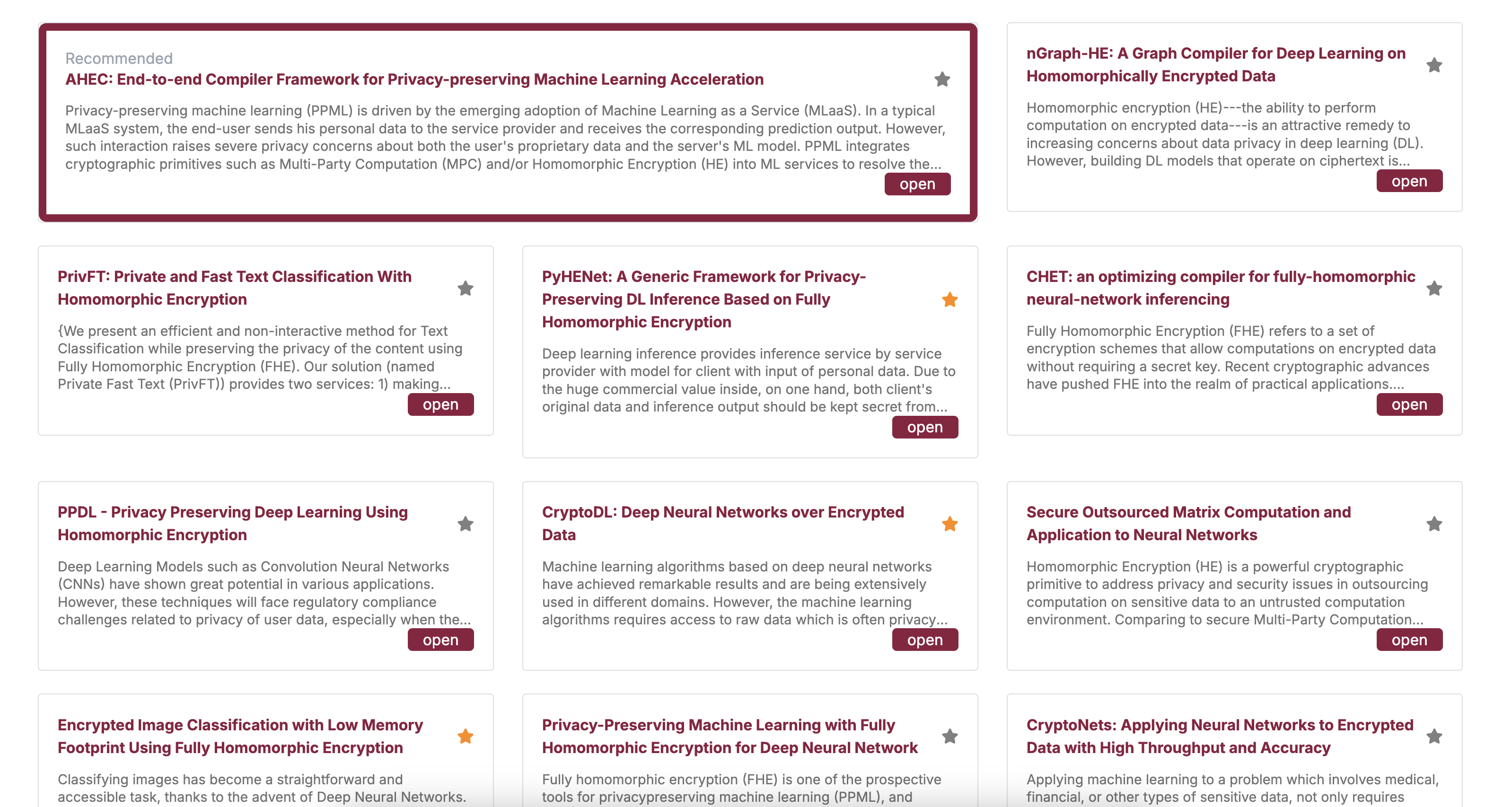}
    \caption{\Prismo Search results}
    \label{fig:appendixsearch}
\end{figure}

\begin{figure}[http]
    \vspace{0.2in}
    \centering
    \includegraphics[width=1.0\linewidth]{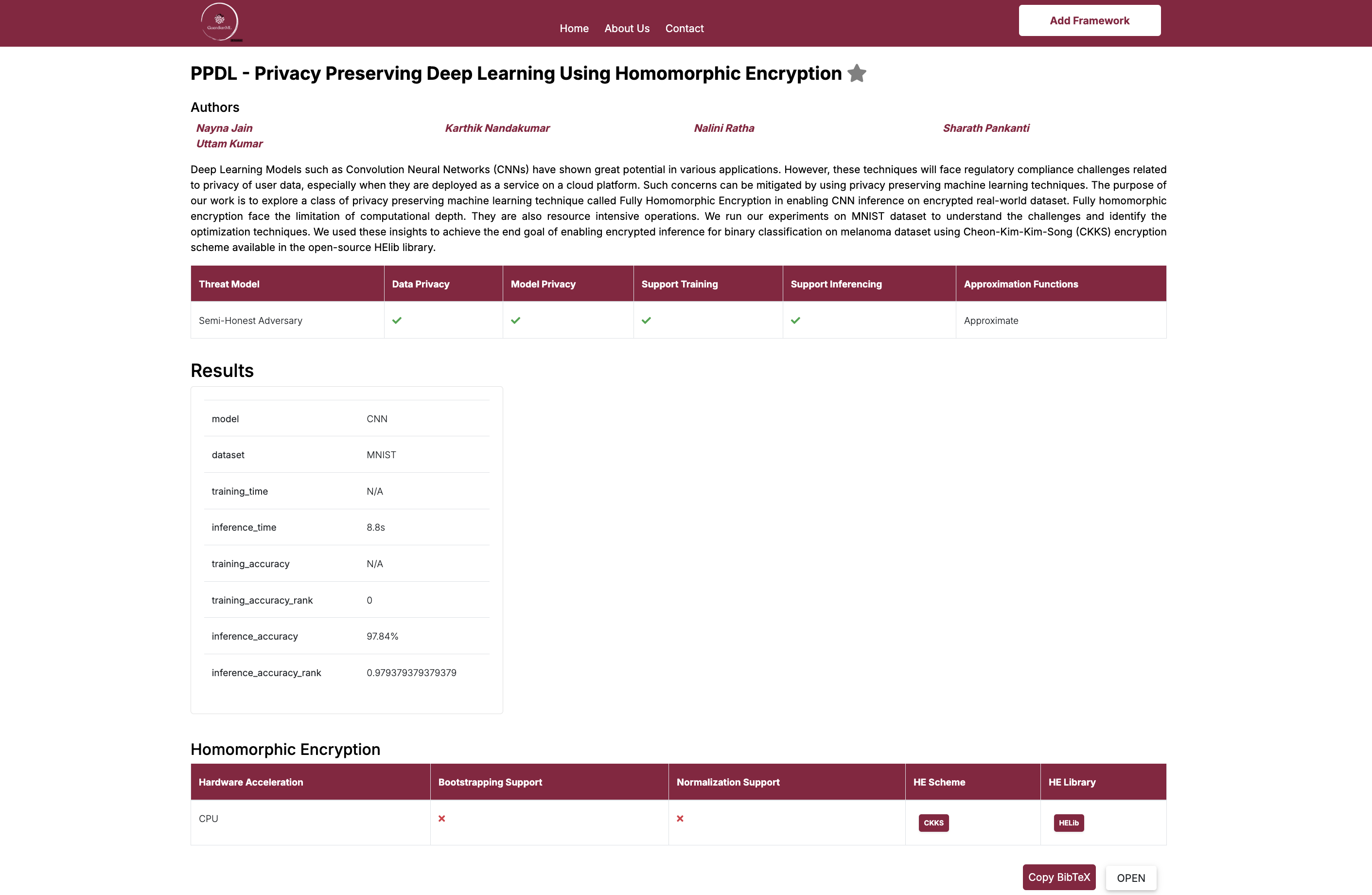}
    \caption{\Prismo Inner Page}
    \label{fig:appendixinner}
\end{figure}



\end{document}